\documentclass[article]{IEEEtran}
\usepackage[cmex10]{amsmath}
\usepackage[short]{optidef}
\usepackage{color}
\usepackage{commath}
\usepackage{algorithm}
\usepackage{algorithmic}
\usepackage{textcomp}
\usepackage{gensymb}
\usepackage{multirow}
\usepackage{multicol}
\usepackage{mathtools}
\ifCLASSINFOpdf
   \usepackage[pdftex]{graphicx}
\graphicspath{ {figures/} }
\else
\fi
\usepackage{graphicx}
\usepackage{stfloats}
\usepackage[font=footnotesize,caption=false,subrefformat=parens,labelformat=parens]{subfig}

\usepackage{cite}
\usepackage{amssymb}
\usepackage{amsthm}
\usepackage{url}
\usepackage{pifont}
\newcommand{\cmark}{\ding{51}}%
\newcommand{\xmark}{\ding{55}}%

\usepackage{siunitx}

\usepackage{eucal}
\usepackage{enumitem}

\newcommand\blfootnote[1]{%
		\begingroup
		\renewcommand\thefootnote{}\footnote{#1}%
		\addtocounter{footnote}{-1}%
		\endgroup
}

\title{Ensuring Reliable Connectivity to Cellular-connected UAVs with Up-tilted Antennas and Interference Coordination}

\author{Md Moin Uddin Chowdhury,
\.{I}smail G\"{u}ven\c{c}, Walid Saad, and Arupjyoti Bhuyan
\thanks{M.M.U. Chowdhury and \.{I}. G\"{u}ven\c{c} are with the Department of Electrical and Computer Engineering, North Carolina State University, Raleigh, NC 27606 (e-mail:~\{mchowdh,iguvenc\}@ncsu.edu).}
\thanks{W. Saad is with the Wireless@VT, Electrical and Computer Engineering Department, Virginia Tech, VA 24060 (e-mail:~walids@vt.edu).}
\thanks{A. Bhuyan is with the Idaho National Laboratory (INL), Idaho Falls,
ID 83402 (e-mail:~arupjyoti.bhuyan@inl.gov).}}

\begin{document}
\pdfoutput=1
\maketitle
\begin{abstract}
To integrate unmanned aerial vehicles (UAVs) in future large-scale deployments, a new wireless communication paradigm, namely, the cellular-connected UAV has recently attracted interest. However, the line-of-sight dominant air-to-ground channels along with the antenna pattern of the cellular ground base stations (GBSs) introduce critical interference issues in cellular-connected UAV communications. In particular, the complex antenna pattern and the ground reflection (GR) from the down-tilted antennas create both coverage holes and patchy coverage for the UAVs in the sky, which leads to unreliable connectivity from the underlying cellular network. To overcome these challenges, in this paper, we propose a new cellular architecture that employs an extra set of co-channel antennas oriented towards the sky to support UAVs on top of the existing down-tilted antennas for ground user equipment (GUE). To model the GR stemming from the down-tilted antennas, we propose a path-loss model, which takes both antenna radiation pattern and configuration into account. Next, we formulate an optimization problem to maximize the minimum signal-to-interference ratio (SIR) of the UAVs by tuning the up-tilt (UT) angles of the up-tilted antennas. Since this is an NP-hard problem, we propose a genetic algorithm (GA) based heuristic method to optimize the UT angles of these antennas. After obtaining the optimal UT angles, we integrate the 3GPP Release-10 specified enhanced inter-cell interference coordination (eICIC) to reduce the interference stemming from the down-tilted antennas. Our simulation results based on the hexagonal cell layout show that the proposed interference mitigation method can ensure higher minimum SIRs for the UAVs over baseline methods while creating minimal impact on the SIR of GUEs.
\end{abstract}

\section{Introduction}
\blfootnote{This work has been supported by NSF grants CNS-1453678, CNS-1910153, CNS-1909372, as well as by Idaho National Laboratory Directed Research Development (LDRD) Program under DOE Idaho Operations Office Contract DEAC07-05ID14517. } 
As the development of the fifth-generation (5G) and beyond wireless networks is underway, unmanned aerial vehicles (UAVs) are expected to play an instrumental role in improving the network capacity and efficiency~\cite{geraci_2018, lin_sky,rui1,zeng2017energy}. While UAVs were originally developed for military applications, due to their fluid mobility, line-of-sight (LOS) transmission, and steadily decreasing production costs, UAVs have been widely used in various new civilian applications, such as packet delivery, search and rescue, video surveillance, aerial photography, airborne communications, among others~\cite{Ramy_walid_ICC2020,Moin-3d, moin3,ali}. 

However, most commercial UAVs acting as aerial users are still dependent on the instructions/maneuvers sent to them by their associated ground pilots through simple direct point-to-point communications. More specifically, this, in turn, limits the UAV use cases to the visual or radio LOS range only. Thus, to take full advantage of large-scale UAV deployment, beyond visual line of sight (BVLOS) UAV operations are of critical importance where the UAVs can reliably obtain command and control (C\&C) communication in the downlink (DL) for safe autonomous operations. In light of such requirements, existing cellular networks can be a strong candidate for deploying autonomous UAVs in BVLOS scenarios with their widespread footprints~\cite{geraci2021,lin_sky}. In fact, field trials from separate industrial entities reported that the existing long-term evolution (LTE) network is capable of meeting some basic requirements of UAV-ground communications~\cite{lin_field,lin_sky}. However, these studies and the Third Generation Partnership Project (3GPP) also pointed out several challenges such as strong inter-cell interference and service of UAVs through antenna side lobes, among others. These challenges come into play due to the fact that traditional cellular networks are optimized for ground user equipment (GUE) by tilting the main lobe of the antennas towards the GUEs. Hence, UAVs flying in the sky are only served by the upper antenna side lobes and experience abrupt signal fluctuations as the UAVs change their locations. Moreover, UAVs also obtain more frequent LOS channels than GUEs. This results in severe interference in the DL from the nearby ground base stations (GBSs) to the UAVs. 

The down-tilted antennas of the existing GBSs can also create another source of interference for the UAVs through the reflected signal from the down-tilted antennas~\cite{wietfeld_ground_reflection}. The main lobe of the antenna hits the ground with an incident angle and the reflected signal can cause non-trivial interference to the UAVs flying in the sky. The non-trivial impact of ground reflection (GR) at millimeter-wave (mmWave) bands is also discussed in~\cite{simran_GR,arup}, where authors introduce the concept of co-channel up-tilted and down-tilted antennas for serving UAVs and GUEs in the mmWave domain. Their ray-tracing-based simulations captured the impact of the angular separations between these two antennas on the coverage performance of the network. However, the authors did not consider the presence of multiple GBSs in their work. The presence of separate co-channel up-tilted antenna sets can help network providers to ensure a high signal-to-interference ratio (SIR) for cellular-connected UAVs. However, proper adjustment of the up-tilt (UT) angles is of critical importance since these extra antennas can create strong LOS interference towards the UAV-GBS links of the network~\cite{simran_GR}. The works in~\cite{geraci_2018,lin_sky} also suggested such dedicated up-tilted cells for serving the UAVs; however, to the best of our knowledge, no prior work considers the problem of tuning the up-tilted antennas for obtaining better UAV SIR performance in a multi-GBS scenario.
\begin{table*}[t]
\centering
\caption{{Literature review.}} 
\scalebox{0.92}{
\begin{tabular}{p{1cm} p{4cm} p{3cm} p{2cm} p{2cm} p{1cm} p{2cm} } \hline
{\textbf{Ref.}} & {\textbf{Goal}} & {\textbf{Interference mitigation technique}} & {\textbf{Antenna radiation pattern}} & {\textbf{up-tilted antenna}} &{\textbf{GR}} & {\textbf{Co-channel UAV \& GUE}}  \\ \hline

\cite{Ramy_walid_ICC2020} & Performance analysis of UAVs considering $3$D antenna radiation & \xmark & directional, array & \xmark & \xmark & \xmark \\ \hline

\cite{ramy_coverage} & Provide reliable connectivity and mobility support for UAVs & Cooperative transmission among GBSs & directional, array & \xmark & \xmark & \xmark \\ \hline

\cite{ramy_antenna} & Simultaneous content delivery to GUEs and UAVs  & MIMO conjugate beamforming & directional, array & \xmark & \cmark \\ \hline

\cite{Mei_interference} & Mitigate
the strong downlink interference to UAVs & Cooperative beamforming & directional, array & \xmark & \xmark & \xmark \\ \hline

\cite{galkin2020} & Intelligent GBS association for UAVs based on network information & Choosing the best GBS by supervised learning & directional, array & \xmark & \xmark & \xmark \\ \hline

\cite{abhay_rws} & Maximize the coverage probability and fifth-percentile rate in hetnet & Optimizing UAV-BS locations and ICIC parameters using exhaustive search & directional, single & \xmark & \xmark & \cmark \\ \hline

\cite{mahdiAari_RL2020} & To reduce disconnectivity time, handover rate, and energy consumption of UAV & Finding the optimal UAV velocity by RL & directional, array & \xmark & \xmark & \xmark \\ \hline

\cite{moin_ICC} & Serve both GUEs
and UAVs simultaneously in a co-channel sub-6 GHz network & Finding the ideal tilting angle by RL & directional, array & \xmark & \xmark & \cmark \\ \hline

\cite{xingqin_rl_2019} & To ensure robust wireless
connectivity and mobility support for UAVs & NA & directional, array & \xmark & \xmark & \xmark \\ \hline

\cite{lin2020a2g} & Maximize aircraft user throughput by tuning ISD and UT angles & Bidirectional deep learning & directional, array & \cmark & \xmark & \xmark \\ \hline

\cite{simran_GR} & Serve both GUEs
and UAVs simultaneously in a co-channel mmWave network & Finding the ideal tilting angle of a single GBS by ray-tracing & directional, single & \cmark & \cmark & \cmark \\ \hline

This work & Maximize the minimum UAV SIR & Tuning the UT angles by GA & directional, array & \cmark & \cmark& \cmark \\ \hline
\end{tabular}}
\label{tab:lit_review}
\end{table*}

Note that, in such a two-antenna setup, the down-tilted antennas create interference to the UAVs by antenna side lobes and the GR. Moreover, the down-tilt (DT) angles of the down-tilted antennas can impact the DL performance of the GUEs as they can be tuned to mitigate the inter-GBS interference for GUEs. Hence, it may not always be possible or convenient to tune the DT angles of cellular networks to optimize coverage for both ground and aerial users. Thus, to mitigate the interference stemming from the down-tilted antennas on the UAVs, we can consider existing inter-cell interference coordination (ICIC) techniques already developed for heterogeneous networks, namely, the 3GPP Release-10 specified enhanced inter-cell interference coordination (eICIC)~\cite{guvenc_letter_eICIC,abhay_rws}. 

Motivated by all these factors, the main contribution of this paper is a novel cellular architecture that leverages additional sets of antennas focusing towards the sky to support UAVs along with existing down-tilted antennas for GUEs. Our key contributions can be summarized as follows:

\begin{itemize}
   \item We first introduce and study a new cellular concept to increase the coverage of cellular-connected UAVs. As mentioned earlier, we propose to use extra antennas with UT angles installed on top of the existing down-tilted antennas for the GUEs. To the best of our knowledge, there are only limited studies in the literature for such an architecture~\cite{simran_GR,arup}. The antenna sets use the same time and frequency resources as the existing down-tilted antennas. However, they focus their main beams towards the sky to provide a more efficient and reliable connectivity to the UAVs.
    
    \item Unlike other previous works, in our proposed architecture, we also consider the presence of GR stemming from the down-tilted antennas while considering the antenna radiation pattern of the down-tilted antennas. To represent the impact of antenna directivity, we modify the GR-based path-loss model introduced in \cite{wietfeld_ground_reflection} to capture the impact of the antenna directivity. Depending on the DT angles of the down-tilted antennas, our analysis shows that the GR can create stronger interference than the antenna's side lobes when the horizontal distance between the UAV and a GBS increases.
    
    \item By considering an interference-limited DL cellular network, we formulate an optimization problem to maximize the minimum SIR of the UAVs by tuning the UT angles of all the up-tilted antennas in the network. Since this is an NP-hard problem, we propose a simple meta-heuristics-based technique, which tunes the UT angles of the GBSs to ensure high minimum UAV SIR. Our proposed method uses the genetic algorithm (GA), a well-known meta-heuristics algorithm that can generate suboptimal solutions efficiently in an iterative method ~\cite{GA_book}.
    
    \item Since the UAVs will experience interference from the extra up-tilted antenna sets along with the antenna side lobes and GRs of the down-tilted antennas, here, we consider the 3GPP Release-10 specified eICIC technique to ensure the reliable coexistence of cellular-connected UAVs and GUEs. The basic idea is that the down-tilted antennas will stop transmission during some portions of the data transmission duration to reduce interference at the UAVs in DL. We discuss eICIC briefly later in this paper.
    
    \item We conduct and present extensive simulations to study the minimum SIR performance of our proposed method. We first obtain suboptimal solutions from the proposed GA-based technique and then use eICIC to further increase the SIR. Our results show that it is possible to obtain high signal-to-interference (SIR) at the UAVs' end by optimizing the UT angles along with considering the eICIC method. By considering different UAV heights and inter-GBS distances, we also show the effectiveness and superiority of our method over some baseline methods. Our results also revealed some interesting yet important design guidelines such as the impact of the number of antenna elements and the DT angles while considering the coexistence of UAVs and GUEs.
    
\end{itemize}

The rest of the paper is organized as follows. We provide a literature review related to the interference mitigation techniques for cellular-connected UAV in Section~\ref{lit_review}. In Section~\ref{sec:sys}, we describe our system model. Section~\ref{sec:max_uptilt_angle} discusses the UT angle maximization problem. We discuss our proposed GA-based UT antenna optimization method in Section~\ref{sec:GA}. Simulation results and the pertinent discussions are presented in Section~\ref{sec:simu_rl}. Finally, conclusions are drawn in Section~\ref{sec:Conc}. The notation list of this paper is presented in Table~II.
\begin{table*}[t]
\begin{center}
\caption{{Notation List.}} 
\scalebox{0.92}{
 {\begin{tabular}{|c||c|c||c|}
  \hline
  \textbf{Notation} & \textbf{Description}  \\ 
   \hline  
 $h_{\rm UAV}$ & UAV altitude  \\ \hline  
  $P_{\rm GBS}$ & Transmit power of the GBSs \\ \hline
  $\mathcal{A}$ & Set of UAV locations \\ \hline
 $\mathcal{B}$ & Set of GBS \\ \hline
  $N_t$ & Number of vertically placed antennas  \\ \hline
  $\phi_{\rm u}$ & Up-tilt angle of the up-tilted antennas  \\ \hline
  $\phi_{\rm d}$ & Down-tilt angle of down-tilted antennas  \\ \hline
  $h_{\textrm{GBS}}^{\textrm{(u)}}$ &  Height of the up-tilted antennas\\ \hline
  $h_{\textrm{GBS}}^{\textrm{(d)}}$ &  Height of the down-tilted antennas\\ \hline
  $h_{\rm d}$ & Height difference between up-tilted and down-tilted antennas\\ \hline
  $\theta_{\rm d}$ & Elevation angle w.r.t. down-tilted antennas\\ \hline
  $G_e(\theta_{\rm d})$ & Element gain w.r.t. down-tilted antennas\\ \hline
  $G_e^{\textrm{max}}$ & Maximum gain of each antenna element\\ \hline
  $G^{\rm (d)}(\theta_{\rm d})$ & Total antenna gain at elevation angle $\theta_{\rm d}$ w.r.t. down-tilted antennas \\ \hline
  $G^{\rm (u)}(\theta_{\rm u})$ & Total antenna gain at elevation angle $\theta_{\rm u}$ w.r.t. up-tilted antennas \\ \hline
  $\mathrm{G_m}$ & Side-lobe level limit \\ \hline
  $P_j^{\textrm{(u)}}$ & Received power from the up-tilted antennas of GBS $j$ \\ \hline
  $P_j^{\textrm{(d)}}$ & Received power from the down-tilted antennas of GBS $j$ \\ \hline
  $\lambda$ & Wavelength of the carrier frequency\\ \hline
  $\hat{G}_j^{\textrm{(v)}}(\theta_{\rm v})$ & Height-dependent antenna gain of the direct path\\ \hline
  $\widetilde{G}_j^{\textrm{(d)}}(h)$ & Height-dependent antenna gain of the reflected path  \\ \hline
  $\psi_j$ & Angle of reflection of GBS $j$\\ \hline
  $R(\psi_j)$ & Ground reflection coefficient for the angle of reflection $\psi_j$ of GBS $j$ \\ \hline
  $\Delta \phi_j$ & Phase difference between the reflected and the direct signal paths of GBS $j$ \\ \hline
  $\alpha(h)$ & UAV height dependent propagation coefficient \\ \hline
  $\hat{G}_j^{\textrm{(d)}}(\psi_j)$ & Antenna gain of the incident path on the ground \\ \hline
  $\gamma_{j,\textrm{usf}}^{\rm (u)}$ & SIR of a UAV connected to up-tilted antennas of GBS $j$ during uncoordinated subframes\\ \hline
  
  $\gamma_{j,\textrm{csf}}^{\rm (u)}$ & SIR of a UAV connected to up-tilted antennas of GBS $j$ during coordinated subframes\\ \hline
  
  $\gamma_{j,\textrm{usf}}^{\rm (d)}$ & SIR of a UAV connected to down-tilted antennas of GBS $j$ during uncoordinated subframes\\ \hline
  
  $\gamma_{j,\textrm{csf}}^{\rm (d)}$ & SIR of a UAV connected to down-tilted antennas of GBS $j$ during coordinated subframes\\ \hline
  
 \end{tabular}}}
\end{center}
\label{tab:notations}
\end{table*}

\label{sec:intro}
\section{Related Works}
\label{lit_review}
Research efforts in integrating UAVs into existing cellular networks with GUEs have recently attracted substantial attention from both academia and industry. For instance, in~\cite{Ramy_walid_ICC2020}, the authors explored the impact of practical antenna configurations on the mobility of cellular-connected UAVs and showed that increasing the number of antenna elements can increase the number of handovers (HOs) for vertically-mobile UAVs. The work in~\cite{mozaffari_tuts} discusses the possibility of using UAVs in wireless networks, with the role of flying base stations and relay nodes.

In~\cite{ramy_coverage}, the same authors provided the upper and lower bounds on the coverage probability of UAVs considering a coordinated multi-point technique. The work in ~\cite{ramy_saad} presented an analytical framework for a coexisting UAV and GUE considering a beamforming technique. By conducting extensive 3GPP compliant simulations, in \cite{xingqin_ho_2018}, the authors showed that the existing cellular networks will be able to support a small number of UAVs with good mobility support. In~\cite{Mozaffari2021Towards6W}, authors summarized the key barriers and their potential solutions for widespread commercial deployment of flying UAVs in beyond 5G wireless systems. Authors in~\cite{alouini2}, proposed an optimization method for managing the movement, charging, and service coverage actions of a fleet of UAVs used as flying base stations. By considering a network of UAV base stations (BSs), the work in~\cite{banagar_ho_wcl2020} introduced exact HO probability for similar UAV velocity and provided lower bound for UAV BSs with different velocities. The authors in~\cite{HO_Ekram_2020} extended the results of~\cite{banagar_ho_wcl2020} by providing exact analysis of HO rate and sojourn time for different UAV velocities and showed that HO rate is minimum when UAV BSs move with the same velocity. However, both of these works treated UAVs as BSs. By using tools from stochastic geometry, the authors in~\cite{Banegar_UT_TWC} studied the performance of 3D two-hop cellular networks where UAV-BSs can obtain wireless backhaul from GBSs. In particular~\cite{Banegar_UT_TWC} considered realistic antenna patterns and dedicated up-tilted antennas for providing better connectivity in the UAV-to-GBS links. 

\begin{figure}[t]
\centering{\includegraphics[width=\linewidth]{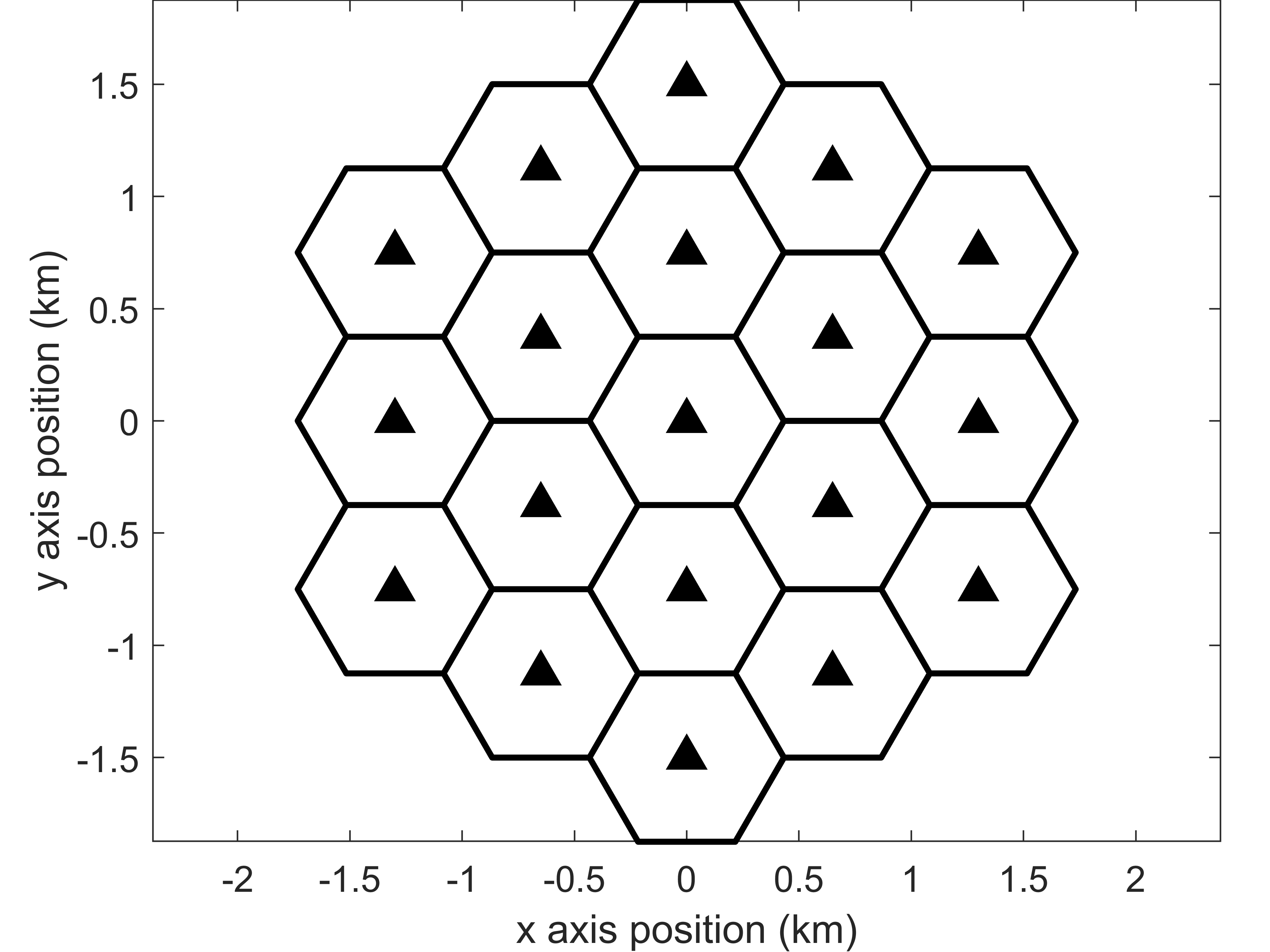}}
    \caption{2-tier hexagonal cell structure with 19 cells and ISD = $500$~m. In this paper, we focus on the center cell with GBS location [0,0] km.}
    \label{fig:cell_structure}
\label{fig:hexa_network}
\end{figure}
Due to the complex antenna pattern and air-to-ground path-loss model, the researcher also relied on learning-based frameworks for ensuring reliable integration and operation of cellular-connected UAVs. For instance, a supervised learning-based association scheme for UAVs was proposed in \cite{galkin2020} to associate UAVs with the GBS providing the highest directional antenna SIR. By tuning the DT angles of the GBSs, the work in~\cite{moin_ICC} used reinforcement learning (RL) to provide good connectivity to both UAVs and GUEs. However, they did not consider the SIR at the UAV which plays a critical role in reliable autonomous UAV deployment. In another work~\cite{galkin2020reqiba}, the authors proposed a deep-learning-based GBS association algorithm for cellular-connected UAVs which takes the knowledge of the cellular environment into account. In the recent work in~\cite{mahdiAari_RL2020}, authors study the problem of jointly optimizing the UAV HO rate, disconnectivity time, UAV flight duration, and UAV energy consumption by tuning the UAV velocity. In particular this prior work explored a multi-armed bandit RL algorithm to solve the problem and showed that the perfect parameters can significantly improve the performance of cellular-connected UAVs. In~\cite{xingqin_rl_2019}, the authors explored an RL algorithm to maximize the received signal quality at a cellular-connected UAV while minimizing the number of HOs. An extension of the traditional RL algorithms known as multi-agent RL has been also introduced for efficient UAV control in~\cite{MFG_MARL_UAV}. Note that these learning-based algorithms will either require advanced data collection, preprocessing, and training, or sample inefficient repetitive interaction with the cellular networks, which makes the deployment of these algorithms challenging for real-world network operators.

In addition to these learning-based methods, non-linear optimization techniques were also used to provide reliable connectivity to UAVs. For instance, in~\cite{meiQF}, the authors proposed a cooperative interference mitigation scheme to mitigate the strong uplink interference from the UAV to a large number of co-channel GBSs serving terrestrial UEs. The helping GBSs sense the UAV's power, which is sent to the main GBS for further interference processing. Similar authors introduced a  cooperative beamforming and transmission scheme to mitigate the interference of cellular-connected UAVs in DL~\cite{Mei_interference}. In~\cite{Mei_UL_NOMA}, they proposed a cooperative non-orthogonal multiple access (NOMA) technique to the uplink communication from a UAV to cellular GBSs, under spectrum sharing with the existing GUEs. The work in~\cite{6G} discusses how to integrate UAVs for providing wireless communications in zones where the deployment of canonical base stations is not possible. In~\cite{jbuh_multi_UAV}, authors introduced the problem of maximizing the minimum UAV rate by joint beamforming, association, and UAV-height control framework for cellular-connected multi-UAV scenarios. However, none of these analytical and learning-based works~\cite{Ramy_walid_ICC2020,ramy_coverage,ramy_saad,xingqin_ho_2018,Mozaffari2021Towards6W,banagar_ho_wcl2020,HO_Ekram_2020,galkin2020,moin_ICC,galkin2020reqiba,mahdiAari_RL2020,xingqin_rl_2019,jbuh_multi_UAV} considered the presence of GR which plays a critical role in air-to-ground communications as an important source of interference for UAVs~\cite{simran_GR,wietfeld_ground_reflection}.
\begin{figure*}[t]
\centering{\includegraphics[width=0.8\linewidth]{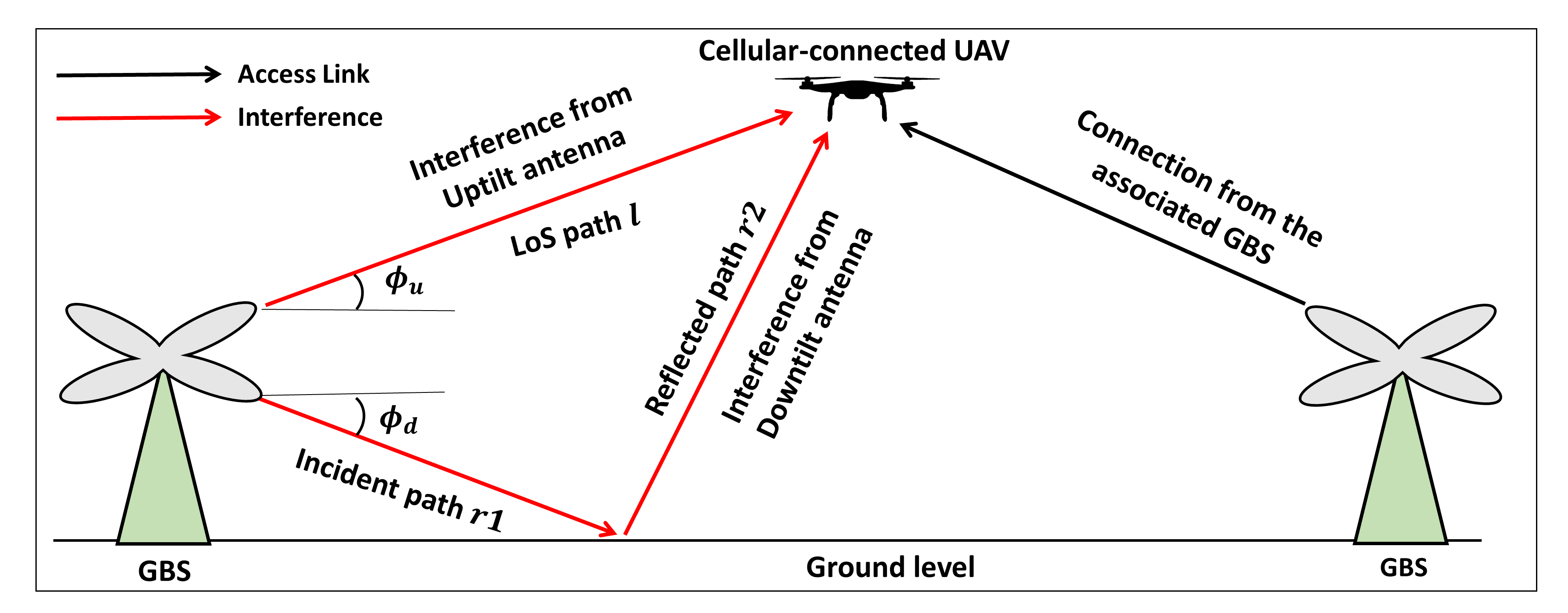}}
    \caption{Illustration of the inter-cell interference at a cellular-connected UAV from the GR signal of a downtilted antenna and the LOS signal from the uptilted antenna of a nearby base station. Though not shown in the figure, the associated GBS in the right can also create interference by the downtilted antennas. The signal quality at the UAV will be effected by the UT angles of the uptilted antennas since they will impact both the desired and the interference signals.}
    \label{fig:network_ground_reflection}
\end{figure*}
The most closely related work here is~\cite{lin2020a2g}, in which the authors introduced a bidirectional deep learning-based technique to maximize the median capacity of an aircraft flying at a height of $12$ km. Using system-level simulation, they considered optimizing the inter-GBS distance and dedicated up-tilted antennas to solve network optimization problems. In contrast to their work, here, we focus on the UAVs flying under $400$ meters of height where the impact of GR is not negligible. Moreover, in our considered system, each GBS can individually change its UT angle, in contrast to the similar UT angles that are assumed for all GBSs in~\cite{lin2020a2g}. To further increase the minimum SIR, we consider the concept of the eICIC to mitigate the interference stemming from the down-tilted antennas at the UAV's end. Since eICIC was already studied extensively in the last decade for increasing efficiency and capacity of the heterogeneous networks~\cite{guvenc_letter_eICIC,abhay_rws}, it will be practical to deploy it for mitigating the interference from the down-tilted antennas. Moreover, the UT angle tuning is based on the GA algorithm, which is also well-studied and was used extensively in optimizations of different aspects of wireless networks~\cite{GA_in_wireless}. For convenience, we summarize and compare the state of the art in the literature with our work in Table~\ref{tab:lit_review}.

\section{System Model}
\label{sec:sys}
\subsection{Network Model}
We consider an interference-limited DL transmission scenario from terrestrial GBSs to cellular-connected UAVs where the $19$ GBSs are distributed in a two-tier hexagonal grid with a fixed inter-site distance (ISD). An illustration of such a network is presented in Fig.~\ref{fig:hexa_network}. Here, we do not consider wraparound~\cite{3gpp.38.901,3gpp} and thus, we will only focus on the performance of the central hexagonal cell to capture the impact of inter-cell interference from the neighboring cells. However, our analysis can easily be extended to larger cellular networks with different GBS distributions. Hereinafter, we will use the terms `GBS' and `cell' interchangeably. To average out the impact of UAV distribution, we divide the center cell into discrete grid points, and a UAV is placed on each grid point at a height $h_{\textrm{UAV}}$. Note that a closer inter-UAV distance or higher grid resolution will provide more fine-grained information on the cellular network characteristics such as interference, GBS association, received signal strength, etc. at the height $h_{\textrm{UAV}}$. Each UAV is assumed to be equipped with a single omnidirectional antenna. The set of the UAV locations and the GBSs can be expressed as $\mathcal{A}$ and $\mathcal{B}$, respectively.

We also assume that all GBSs have equal altitudes $h_{\textrm{GBS}}$ and transmission power $P_{\textrm{GBS}}$. The GBSs consist of $N_t$ vertically placed cross-polarized directional antennas down-tilted by angle $\phi_{\rm d}$~\cite{Ramy_walid_ICC2020,Moin-3d}. We consider the GBS antennas to be omnidirectional in the horizontal plane but they have a variable radiation patterns along the vertical dimension with respect to the elevation angle between the antennas and the users~\cite{ramy_antenna}. 

Different from the traditional cellular network setting, here, we also consider the presence of another set of antennas on top of the previous ones, which can provide connectivity to the UAVs using UT angle $\phi_{\rm u}$. Since the UAVs served by only down-tilted antennas suffer from poor connectivity and severe interference, up-tilted antennas can be used to provide reliable connectivity to the UAVs~\cite{geraci_2018,simran_GR}. Note that the antenna tilt angle is obtained by introducing a fixed phase shift to the signal of each element. We define $h_{\textrm{GBS}}^{\textrm{(u)}}$ and $h_{\textrm{GBS}}^{\textrm{(d)}}$, respectively, as the height of the up-tilted antennas and down-tilted antennas. The two sets of antenna setups are separated by a height difference $h_{\rm d}$, i.e., $h_{\rm d}=h_{\textrm{GBS}}^{\textrm{(u)}}-h_{\textrm{GBS}}^{\textrm{(d)}}$. We consider that all of the GBSs and their sets of antennas share the same time and frequency resources. The UAVs will be associated with the antenna set (up-tilted or down-tilted) of the GBS providing the highest reference signal received power (RSRP)~\cite{Ramy_walid_ICC2020,lin_mobility}. 

\subsection{Antenna radiation pattern}
The $N_t$ antennas are equally spaced where adjacent elements are separated by half-wavelength distance. The element power gain (in dB) in the vertical plane at elevation angle $\theta_{\rm d}$ with respect to the down-tilted antennas can be specified by~\cite{3gpp.38.901}
\begin{equation}
    G_e(\theta_{\rm d})=G_e^{\textrm{max}}- \text{min}\left\{ 12\left(\frac{\theta_{\rm d}}{\theta_{3\mathrm{dB}}} \right)^2, \mathrm{G_m}\right\},
\end{equation}
where $\theta_{\rm d} \in [-90^\circ, 90^\circ]$, $\theta_{3\textrm{dB}}$ refers to the $3$ dB beam width with a value of $65^\circ$, $G_e^{\textrm{max}}=8$~dBi is the maximum gain of each antenna element, and $\mathrm{G_m}$ is the side-lobe level limit, respectively, with a value $30$ dB~\cite{rebato_antenna}. Note that $\theta_{\rm d}=0^\circ$ refers to the 
horizon and the $\theta_{\rm d}=90^\circ$ represents the case when the main beam is facing upward perpendicular to the $xy$-plane~\cite{3gpp.38.901}. The array factor $A_f^{\rm d}(\theta_{\rm d})$ of the ULA with $N_t$ elements while considering a DT angle $\phi_{{\rm d}}$ is given by
\begin{equation}
    A_f^{\rm (d)}(\theta_{\rm d})=\frac{1}{\sqrt{N_t}}\frac{\sin\big({\frac{N_t\pi}{2}} (\sin\theta_{\rm d}-\sin\phi_{\rm d})\big)}{\sin\big({\frac{\pi}{2}} (\sin\theta_{\rm d}-\sin\phi_{\rm d})\big)}.
\end{equation} 
Let us denote $G_f^{\rm (d)}(\theta_{\rm d})\triangleq 10\log_{10}( A_f^{\rm d}(\theta_{\rm d}))^2 $ as the array power gain in dB scale. Then the overall antenna gain at elevation angle $\theta_{\rm d}$ is given by 
\begin{equation}
    G^{\rm (d)}(\theta_{\rm d})=G_e(\theta_{\rm d})+G_f^{\rm (d)}(\theta_{\rm d}).
\label{eq:total_antenna_gain_down}
\end{equation} 
Similarly, the array factor pertinent to the up-tilted antennas with UT angle $\phi_{\rm u}$ and elevation angle $\theta_{\rm u}$ can be expressed as: \vspace{-0.3cm}
\begin{equation}  \vspace{-0.2cm}
    A_f^{\rm (u)}(\theta_{\rm u})=\frac{1}{\sqrt{N_t}}\frac{\sin\big({\frac{N_t\pi}{2}} (\sin\theta_{\rm u}-\sin\phi_{\rm u})\big)}{\sin\big({\frac{\pi}{2}} (\sin\theta_{\rm u}-\sin\phi_{\rm u})\big)}.
\end{equation}
The array gain $G_f^{\rm (u)}(\theta_{\rm u})\triangleq 10\log_{10}( A_f^{\rm u}(\theta_{\rm u}))^2$ can then be derived and, finally, the overall antenna gain due to the UT angle $\phi_{\rm u}$ can be expressed as: 
\begin{equation}
G^{\rm (u)}(\theta_{\rm u})=G_e(\theta_{\rm u})+G_f^{\rm (u)}(\theta_{\rm u}).
\label{eq:total_antenna_gain_up}
\end{equation}
\subsection{Ground reflection channel model}
The channel between a GBS and a UAV plays a critical role in the coverage performance at the UAV's end and we consider a channel model that is characterized by both distance-based path-loss and GR. To characterize the GR, we modify the height-dependent path-loss model introduced in~\cite{wietfeld_ground_reflection} which is a variant of the two-ray path-loss model~\cite{Goldsmith:2005:WC:993515}. Let the length of the $3$D Cartesian distance from a UAV to a GBS $j$ be $l_j$ and the length of the incident and reflected paths are $r_{1,j}$ and $r_{2,j}$, respectively. For convenience, we discard the subscript from $h\textsubscript{UAV}$ in the following analysis.  Finally, the received power from GBS $j$ at a UAV at height $h$ can be specified as:
\begin{equation}
    P_j^{\textrm{(v)}}=P_{\textrm{GBS}}\bigg[\frac{\lambda}{4\pi}\bigg]^2\bigg| \frac{\hat{G}_j^{\textrm{(v)}}(\theta_{{\rm v},j})}{l_j} + \frac{R(\psi_j)\widetilde{G}_j^{\textrm{(d)}}(h)e^{i\Delta \phi_j}}{r_{1,j}+r_{2,j}}\bigg|^{\alpha(h)},
\label{eq:rx_power_gr}
\end{equation}
where $\textrm{v} \in \{\textrm{u},\textrm{d}\}$, $\theta_{{\rm v},j}$ is the elevation angle with respect to the up-tilted or down-tilted antenna of GBS $j$, $i=\sqrt{-1}$ is the imaginary unit of a complex number, $\lambda$ is the wavelength of the carrier frequency, $\hat{G}_j^{\textrm{(v)}}(\theta_{\rm v})$ and  $\widetilde{G}_j^{\textrm{(d)}}(h)$ represent the height-dependent antenna gain of the direct and reflected path, respectively, $R(\psi_j)$ is the GR coefficient for the angle of reflection $\psi_j$ with respect to the ground plane, $\Delta \phi_j=(r_{1,j}+r_{2,j})-l_j$ is the phase difference between the reflected and the direct signal paths, and $\alpha(h)$ is the height dependent propagation coefficient for UAV height $h$. Here, we do not consider GR from the up-tilted antennas since their main beams are oriented towards the sky.\looseness=-1 

Note that the GR coefficient for cross-polarized antennas can be calculated as $R(\psi_j)=\frac{R_{\rm H}(\psi_j)-R_{\rm V}(\psi_j)}{2}$~\cite{najibi2013physical},~which also depends on the relative ground permittivity $\epsilon_r \approx 15$~\cite{wietfeld_ground_reflection}, reflection coefficients for horizontal linear polarization $R_{\rm H}(\psi_j)$ and vertical linear polarization $R_{\rm V}(\psi_j)$. Moreover, $\hat{G}_j^{\textrm{(v)}}(\theta_{\rm v})$ depends on the instantaneous elevation angle between the GBS and the UAV by~\eqref{eq:total_antenna_gain_down} and~\eqref{eq:total_antenna_gain_up}, whereas $\widetilde{G}_j^{\textrm{(d)}}(h)$ can be expressed as:
\begin{equation}
\label{eq:ground_reflected}
  \resizebox{\hsize}{!}{$\widetilde{G}_j^{\textrm{(d)}}(h) = \left \{
  \begin{aligned}
    &\hat{G}_j^{\textrm{(d)}}(\psi_j), && h<h_t \\
    &\frac{\hat{G}_j^{\textrm{(d)}}(\psi_j)}{2}, && h_t \leq h\leq 2h_t\\
    &\frac{\hat{G}_j^{\textrm{(d)}}(\psi_j)}{2}-\frac{h}{2 h_{t,c}}\cdot(\hat{G}_j^{\textrm{(d)}}(\psi_j)-1), && 2h_t \leq h\leq 500 \\
    &0.5, && h\geq 500 
  \end{aligned} \right.$}
\end{equation} 
where $h_t=2h_{\textrm{GBS}}^{\textrm{(d)}}+2$ and $h_{t,c}=500$~m are threshold heights~\cite{wietfeld_ground_reflection}, and $\hat{G}_j^{\textrm{(d)}}(\psi_j)$ is the antenna gain of the incident path on the ground from the down-tilted antennas which depends on $N_t$. Finally, the height-dependent propagation coefficient can be expressed as:
\begin{equation}
\label{eq:ground_refle}
  \alpha(h)= \left \{
  \begin{aligned}
    &\alpha_0-h\cdot \bigg(\frac{(\alpha_0-2)}{h_{\textrm{GBS}}^{\textrm{(v)}}}\bigg), &&  h<2\cdot h_{\textrm{GBS}}^{\textrm{(v)}}, \\
    &2 && h\geq 2\cdot h_{\textrm{GBS}}^{\textrm{(v)}},
  \end{aligned} \right.
\end{equation} 
where $\alpha_0$ is the maximum possible attenuation coefficient~\cite{wietfeld_ground_reflection}. Here, we do not consider any GR due to the antenna side lobes. From~\eqref{eq:ground_reflected}, we can see that the antenna gain is dependent on the incident angle $\psi_j$, whereas in~\cite{wietfeld_ground_reflection}, the gain of the reflected path is assumed to be constant with respect to $\psi_j$. In Fig.~\ref{fig:network_ground_reflection}, we provide a simple illustration of how a UAV can suffer from interference from GR and antenna side lobes. 
\begin{figure}[t]
\centering{\includegraphics[width=\linewidth]{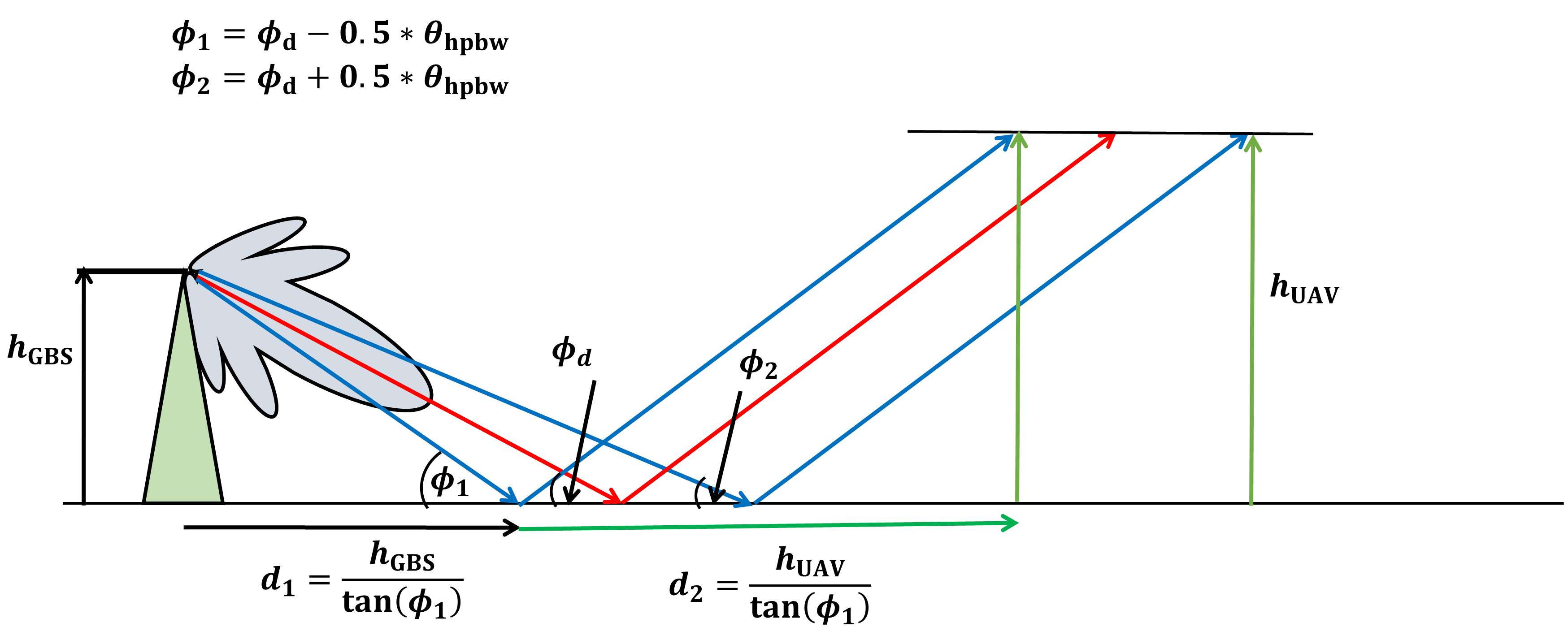}}
    \caption{Analysis of GR depending on the DT angle $\phi_{\rm d}$.}
    \label{fig:ground_reflection}
\end{figure}
\textbf{Remark 1}: \textit{Due to the the DT angle $\phi_{\rm d}$, the main lobe of the down-tilted antenna will not reach the ground level before the horizontal distance (in meter) is away by $\frac{h_{\rm GBS}}{\tan(\phi_{\rm d})}$ from the GBS. Hence, UAVs closer to this distance from a GBS will not be impacted by the GR stemming from the down-tilted main lobe of that particular GBS}.

Next, for a given UAV height and DT angle, we derive the distances from a GBS where the impact of the GR is the most effective.

\textbf{Theorem 1}: \textit{For a given $h_{\rm GBS}$, $h_{\rm UAV}$, and DT angle $\phi_{\rm d}$, the impact of the GR from a GBS will mostly be seen between horizontal distances $d_1=\frac{h_{\rm GBS}+h_{\rm UAV}}{\tan(\phi_1)}$ and $d_2=\frac{h_{\rm GBS}+h_{\rm UAV}}{\tan(\phi_2)}$ from that GBS, 
where
\begin{align}
   \phi_1&=\phi_{\rm d}-0.5\times \theta_{\rm hpbw} \label{eq:phi1},\\
   \phi_2&=\phi_{\rm d}+0.5\times \theta_{\rm hpbw},
\label{eq:phi2}
\end{align}
and $\theta_{\rm hpbw}$ is half power beam width of the main lobe of the down-tilted antenna.}

\textit{Proof}: Consider a scenario with a single GBS with antenna pattern and height are as specified in Section~\ref{sec:sys}. Since GR only stems from the down-tilted antennas, here, we consider that the GBS is only equipped with down-tilted antenna with DT angle $\phi_{\rm d}$. Let us consider the half-power beam width (HPBW) of the main lobe as $\theta_{\rm hpbw}$. Note that the HPBW is inversely proportional to the number of elements in the antenna array~\cite{antenna_gain_beamwidth}. Given the DT angle $\phi_{\rm d}$, the two angles of the two end points of the HPBW will be as expressed in \eqref{eq:phi1} and \eqref{eq:phi2}.

Then the down-tilted main beam will reach the ground and the impact of the HPBW will be within the distances $r1=\frac{h_{\rm GBS}}{\tan(\phi_1)}$ and $r2=\frac{h_{\rm GBS}}{\tan(\phi_2)}$ from the GBS as depicted in Fig.~\ref{fig:ground_reflection}. By assuming regular reflection from the ground, the two rays will reach the UAV height at a distance $d_1=\frac{h_{\rm GBS}+h_{\rm UAV}}{\tan(\phi_1)}$ and $d_2=\frac{h_{\rm GBS}+h_{\rm UAV}}{\tan(\phi_2)}$, respectively from the GBS, which completes the proof.

Theorem 1 provides us the range of distances from a GBS where a UAV will be impacted significantly by GR for a given DT angle $\phi_{\rm d}$. From Theorem 1, we can observe that for a higher $\phi_{\rm d}$, locations closer to the GBSs will be impacted by GR and vice versa.  
\begin{figure}[t]
\centering
		\subfloat[$h_{\textrm{UAV}}=50$~m.]{
			\includegraphics[width=.9\linewidth]{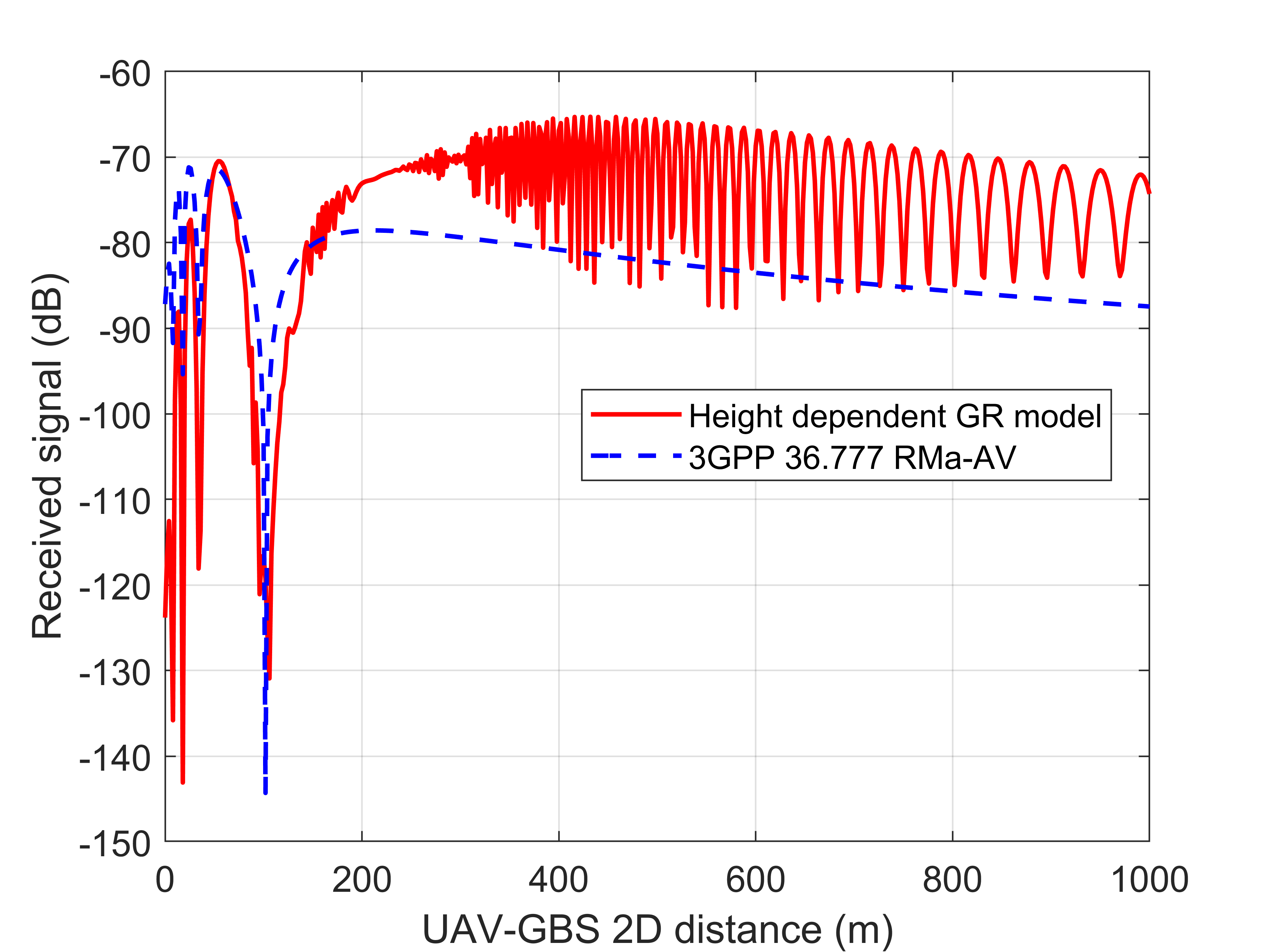}} \hfill
		\subfloat[$h_{\textrm{UAV}}=100$~m.]{
			\includegraphics[width=.9\linewidth]{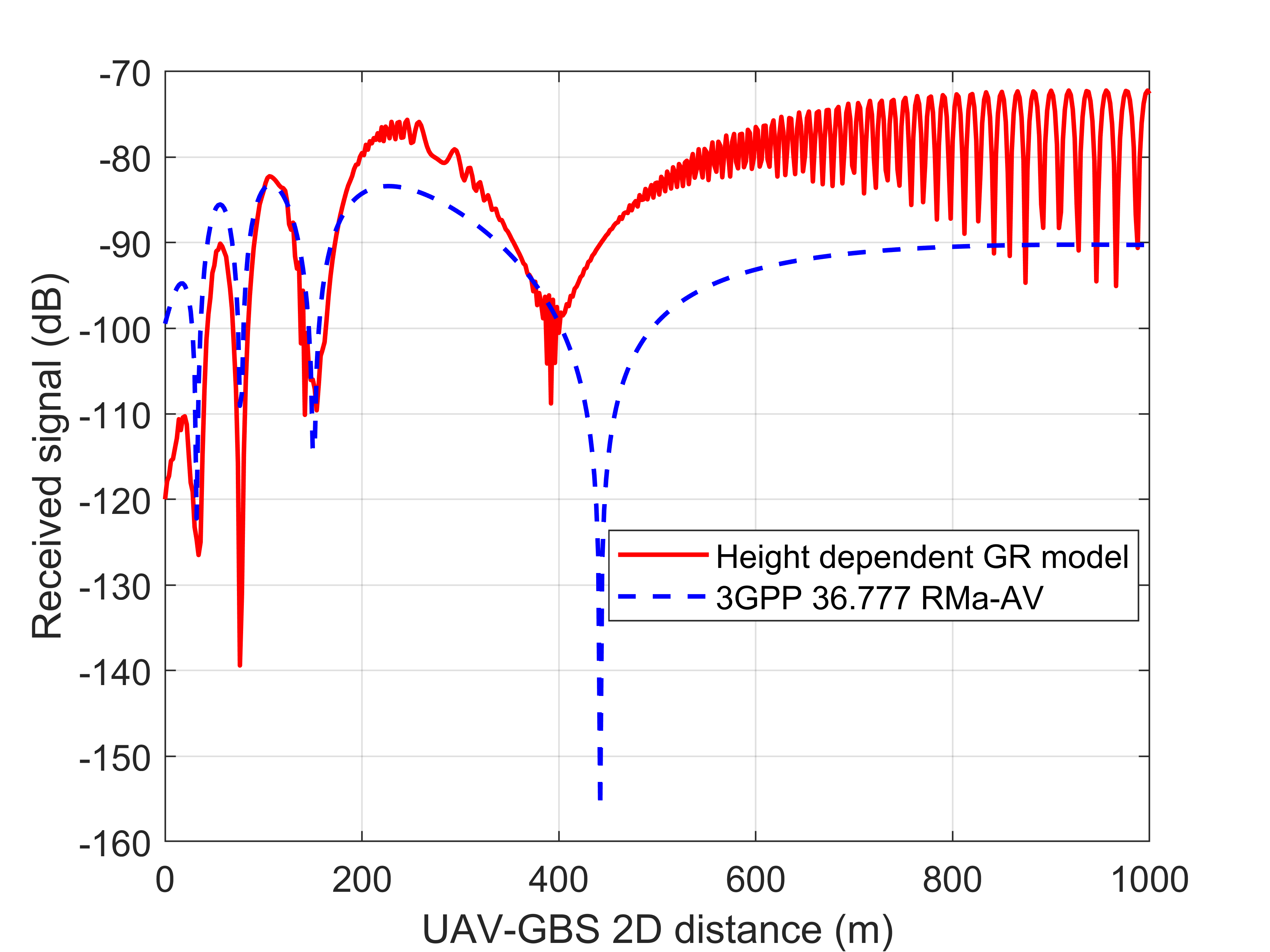}} 
		\caption {
		{Comparison of GR and 3GPP RMa-AV channel model~\cite{3gpp} for different UAV heights considering the antenna radiation pattern and $\phi_{\rm d}=6^\circ$. (a) $h_{\textrm{UAV}}=50$~m } and (b) $h_{\textrm{UAV}}=100$~m.}
		\label{fig:gr_vs_3gpp}
\end{figure}

\textbf{Remark 2}: \textit{If $\phi_{\rm d} < \frac{\theta_{\rm hpbw}}{2}$, then the impact GR at the UAV will start from the distance $d_1$ and will the impact of the main lobe will last till infinity. However, due to the path-loss, the impact will gradually decrease as the horizontal distance increases beyond $d_1$}. 

\subsection{Numerical example}
By considering $\phi_{\rm d}=6^\circ$, in Fig.~\ref{fig:gr_vs_3gpp}(a), we compare the 3GPP RMa-AV model~\cite{3gpp} and our proposed height dependent GR model for $h_{\rm UAV}=50$~m, $h_{\rm GBS}=30$~m, and $P_{\rm GBS}=30$~dBm, while considering the antenna radiation pattern as discussed before. The received signal plot with respect to 2D UAV-BS distance shows that the impact of GR comes into play after a certain horizontal distance. The ripple in the received signal is created due to the phase difference between the direct LoS path and the reflected path and the GR can provide more than $10$ dB more signal power than the 3GPP model. For $h_{\rm UAV}=100$~m, as shown in  Fig.~\ref{fig:gr_vs_3gpp}(b), the GR shows a similar kind of trend but after greater UAV-to-GBS horizontal distance as discussed in Theorem 1. 

\begin{figure}[t]
\centering
		\subfloat[$h_{\textrm{UAV}}=100$~m, $\phi_{\rm d}=6^\circ$.]{
			\includegraphics[width=.9\linewidth]{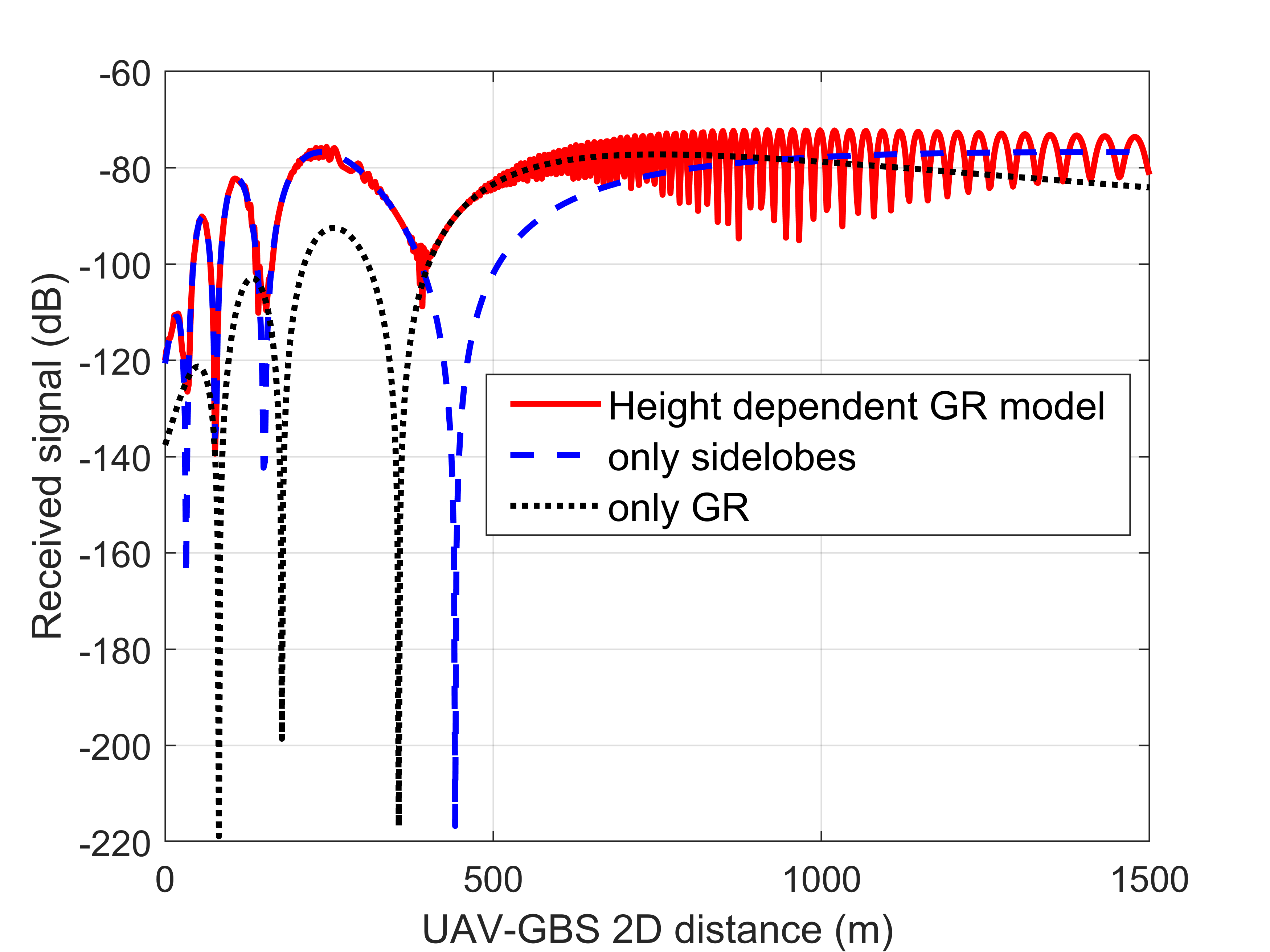}} \hfill
		\subfloat[$h_{\textrm{UAV}}=100$~m, $\phi_{\rm d}=10^\circ$.]{
			\includegraphics[width=.9\linewidth]{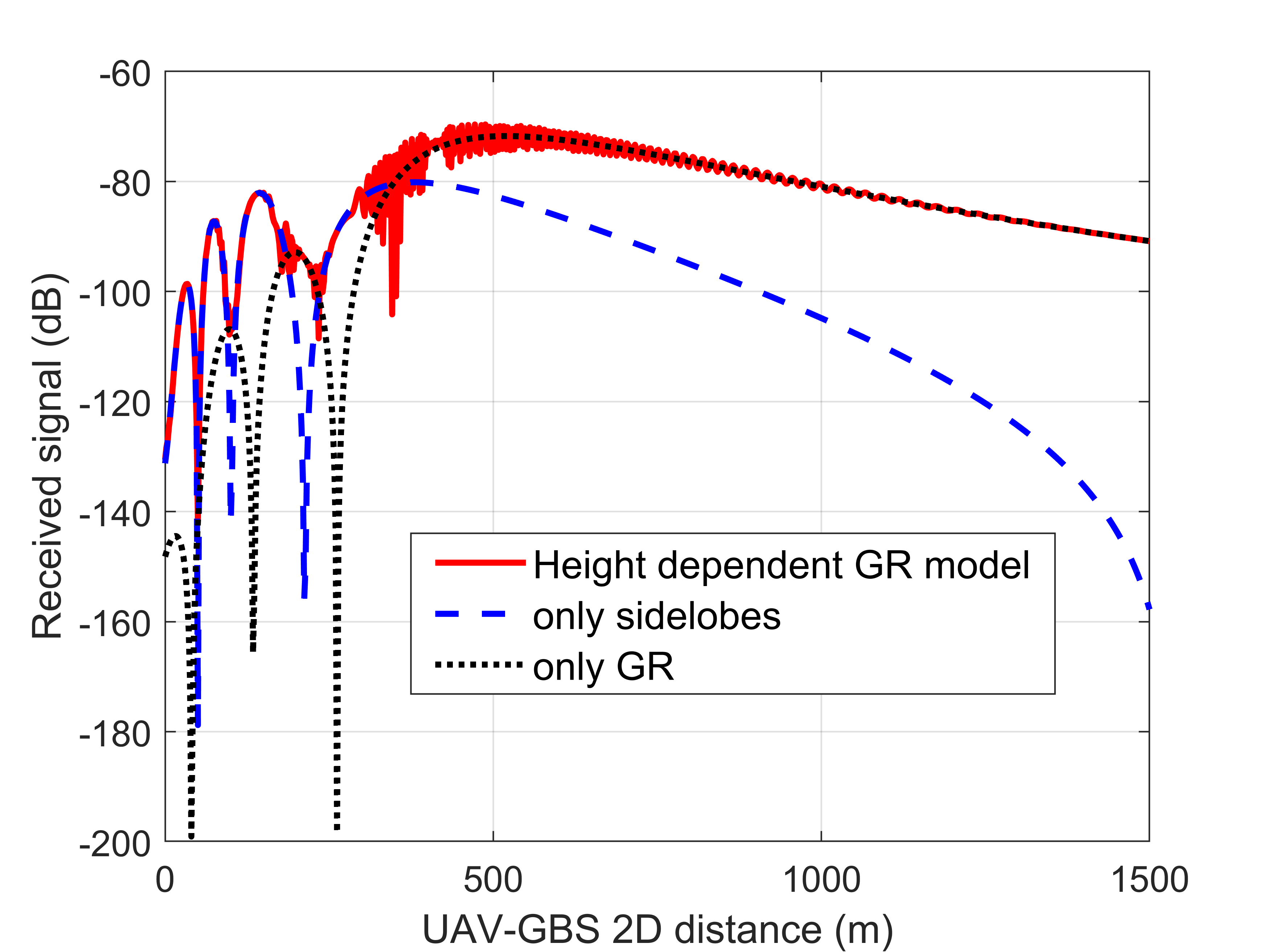}}
		\caption {
		{Impact of GR and antenna side lobes on the GR-based path-loss model for $h_{\textrm{UAV}}=100$~m. (a)  $\phi_{\rm d}=6^\circ$ and (b)  $\phi_{\rm d}=10^\circ$.}}
		\label{fig:gr_vs_dt}
\end{figure}

Finally, we split the reflected signal from the down-tilted antennas into its two ingredients: the signal from the antenna side lobes and the reflected signal from the main beam of the DT antennas. The relevant results for $h_{\rm UAV}=100$~m are shown in Fig.~\ref{fig:gr_vs_dt}(a), from which we conclude that the GR path-loss model coincides with the side lobes when the UAV is close to the GBS. However, after a distance of $400$~m, the GR starts to provide high power through the main lobe which even compensates the antenna's side-lobe null at $442$~m. Overall, the GR keeps dominating the signal from the DT angles till about $900$~m. We also study the impact of GR for higher DT angles in Fig.~\ref{fig:gr_vs_dt}(b). For a DT angle of $10^\circ$, GR starts dominating the signal power from about $350$~m and can act as the dominant source of interference for a UAV situated at a distance of $1500$ meters.
From the above discussion, we can conclude that the down-tilted antennas can create significant interference towards the far UAVs by GR. However, other than some works, the impact of GR is not considered in the literature. Apart from this, the up-tilted antennas can also create strong interference. However, we can mitigate the interference from the up-tilted antennas by tuning the UT angles properly~\cite{simran_GR}. Hence, to increase the reliability of the cellular-connected UAVs, we consider the eICIC method to reduce the interference from the down-tilted antennas.
\subsection{Overview of eICIC}
\label{sec:eICIC}
To mitigate the interference problems caused by the extra set of antennas, we consider eICIC techniques which have been specified in LTE Release-10 of 3GPP~\cite{3gppeICIC}. The time-domain eICIC technique provides an interference coordination method based on the subframe blanking, known as almost blank subframe (ABS) that does not send any traffic channels and sends mostly control channels with very low power. In our proposed interference mitigation method, the down-tilted antennas will not transmit data while allowing the up-tilted antennas to serve UAVs suffering from high interference during an ABS. Transmissions from the down-tilted antennas are periodically muted during the entire frame duration. The up-tilted antennas can send their data during such an ABS and avoid interference. Note that certain control signals are still required to be transmitted even in the muted subframes to avoid radio link failure~\cite{arvind_ho}. 

\begin{figure}[t]
\centering{\includegraphics[width=\linewidth]{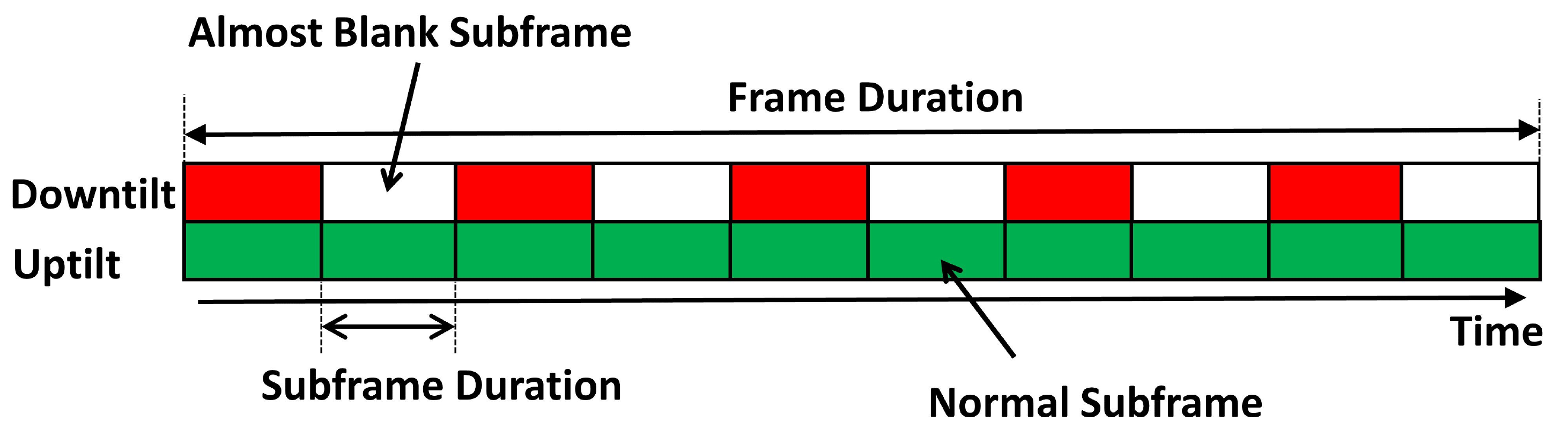}}
    \caption{Basic principle of time domain eICIC. For the considered scenario, the aerial users can be scheduled in the up-tilted antenna subframes that overlap with the almost blank subframes of the down-tilted antennas. This will protect aerial users from the sidelobe interference and the ground reflection interference coming from the down-tilted antennas, as illustrated in Fig.~\ref{fig:network_ground_reflection}.}
    \label{fig:eICIC_architecture}
\end{figure}

The frame structure of the eICIC is shown in Fig.~\ref{fig:eICIC_architecture}. During the uncoordinated subframes (USFs), the down-tilted antennas transmit data and control signals at full power $P_{\rm GBS}$ while during the coordinated subframes (CSFs), they remain muted. We define $\beta$ as the duty cycle of USFs which refers to the ratio of the number of USFs to the total number of subframes in a frame. Then, $(1-\beta)$ will be the duty cycle of the silent subframes or CSFs. Here, we assume full coordination and synchronization among the GBSs and hence, the ABS pattern of all the down-tilted antennas will be the same. We will show in the next subsection that the choice of $\beta$ will impact the capacity/rate of the UAVs/GUEs associated with the down-tilted antennas. However, this is out of the scope of this paper and will be subject of our future work.

\section{Up-tilt Angle Optimization for Maximizing SIR}
\label{sec:max_uptilt_angle}

\subsection{SIR definitions over different subframes}

As mentioned earlier, we consider an interference-limited DL sub-$6$ GHz band for the cellular network, where the presence of thermal noise is omitted. We also assume that the GBSs and both up-tilted and down-tilted antennas share a common transmission bandwidth and full buffer traffic is used in every GBS~\cite{Moin-3d,moin_rws}. Then, we can calculate the SIR of a UAV connected to the up-tilted antennas of GBS $j$ considering flat-fading channels~\cite{guvenc_letter_eICIC} and antenna pattern during USF by the following expression:
\begin{equation}
\label{eq:sir_uptilt}
    \gamma_{j,\textrm{usf}}^{\rm (u)}= \frac{P_{j}^{\rm (u)}}{\sum \limits_{b \in \mathcal{B}, b \neq j,} \sum \limits_{{\rm v} \in \{{\rm u},{\rm d}\}} P_{b}^{\rm (v)}+ P_{j}^{\rm (d)}}.
\end{equation}
Similarly, SIR of a UAV connected to the down-tilt antennas of GBS $j$ considering flat-fading channels during USF as follows:
\begin{equation}
\label{eq:sir_downtilt}
    \gamma_{j,\textrm{usf}}^{\rm (d)}= \frac{P_{j}^{\rm (d)}}{\sum \limits_{b \in \mathcal{B}, b \neq j,} \sum \limits_{{\rm v} \in \{{\rm u},{\rm d}\}} P_{b}^{\rm (v)}+ P_{j}^{\rm (u)}}.
\end{equation}
Note that~\eqref{eq:rx_power_gr} is used to calculate the received power from a particular antenna set (up-tilted/down-tilted) of a GBS. We assume flat-fading channels due to the presence of narrowband OFDM-based communications in existing cellular networks. After considering the antenna radiations from the both sets of antennas and some algebraic calculations, the closed-form expressions of \eqref{eq:sir_uptilt} and \eqref{eq:sir_downtilt} are expressed by \eqref{eq:sir_uptilt_closed} and \eqref{eq:sir_downtilt_closed}, respectively, which are presented on the next page. During the CSFs, the down-tilt antennas are kept off to protect the UAVs from interference (GR of the beam's boresight and the LOS interference from the beam's side lobes). Note that the interference to a UAV served by an up-tilted antenna may be coming also from the down-tilted antenna located at the same GBS. Thus, the SIR of a UAV connected to the up-tilted antennas of GBS $j$ during CSF can be expressed as follows:
\begin{equation}
\label{eq:sir_uptilt_csf}
    \gamma_{j,\textrm{csf}}^{\rm (u)}= \frac{P_{j}^{\rm (u)}}{\sum \limits_{b \in \mathcal{B}, b \neq j}  P_{b}^{\rm (u)}}.
\end{equation}
Finally, we can find the capacity of a UAV connected to up-tilted antennas of GBS $j$ during USFs as follows:
\begin{equation}
{C_{j,\textrm{usf}}^{\rm (u)}}={\log_{2}(1+{ \gamma_{j,\textrm{usf}}^{\rm (u)})}}.
\label{eqn:rate_uptilt}
\end{equation} 
On the other hand, if the UAV is associated with down-tilted antenna of its serving GBS, it will obtain its data in the DL during the USFs. Hence, the rate can be expressed as 
\begin{equation}
{C_{j,\textrm{usf}}^{\rm (d)}}=\beta \big({\log_{2}(1+{ \gamma_{j,\textrm{usf}}^{\rm (d)})}}\big).
\label{eqn:rate_downtilt}
\end{equation}

\begin{figure*}[t]
\hrulefill
\begin{equation} 
\label{eq:sir_uptilt_closed}
\resizebox{\hsize}{!}{ $\gamma_{j,\rm{usf}}^{\rm (u)}= \frac{{10^{\frac{G_e(\theta_{{\rm u},j})}{10}}\Bigg[\frac{\sin^2\big({\frac{N_t\pi}{2}}\big(\sin(\theta_{{\rm u},j})-\sin(\phi_{{\rm u},j})\big)\big) } {\sin^2\big({\frac{\pi}{2}}\big(\sin(\theta_{{\rm u},j})-\sin(\phi_{{\rm u},j})\big)\big)} \big(\frac{1}{l_j} \big) \Bigg]}^{\alpha(h)}}{{\sum \limits_{i \in\mathcal{B}, i \neq j } 10^{\frac{G_e(\theta_{{\rm u},i})}{10}}\Bigg[ \frac{\sin^2\big({\frac{N_t\pi}{2}}\big(\sin(\theta_{{\rm u},i})-\sin(\phi_{{\rm u},i})\big)\big) } {\sin^2\big({\frac{\pi}{2}}\big(\sin(\theta_{{\rm u},i})-\sin(\phi_{{\rm u},i})\big)\big)}\big(\frac{1}{l_i} \big)\Bigg]}^{\alpha(h)}+\frac{1}{N_t}\sum \limits_{k \in\mathcal{B}}\bigg|10^{\frac{G_e(\theta_{{\rm d},k})}{10}}\frac{\sin^2\big({\frac{N_t\pi}{2}}\big(\sin(\theta_{{\rm d},k})-\sin(\phi_{{\rm d},k})\big)\big) } {\sin^2\big({\frac{\pi}{2}}\big(\sin(\theta_{{\rm d},k})-\sin(\phi_{{\rm d},k})\big)\big)}  \big(\frac{1}{l_k} \big) + \frac{R(\psi_k) \widetilde{G}_k^{\textrm{(d)}}(h)e^{k\Delta \phi}}{r_{1,k}+r_{2,k}}\bigg|^{\alpha(h)}}$}.
\end{equation}
\vfill
\begin{equation} 
\label{eq:sir_downtilt_closed}
\resizebox{\hsize}{!}{$\gamma_{j,\rm{usf}}^{\rm (d)}= \frac{{\frac{1}{N_t}\bigg|10^{\frac{G_e(\theta_{{\rm d},j})}{10}}\frac{\sin^2\big({\frac{N_t\pi}{2}}\big(\sin(\theta_{{\rm d},j})-\sin(\phi_{{\rm d},j})\big)\big) } {\sin^2\big({\frac{\pi}{2}}\big(\sin(\theta_{{\rm d},j})-\sin(\phi_{{\rm d},j})\big)\big)} \big(\frac{1}{l_j} \big) + \frac{R(\psi_j)\widetilde{G}_j^{\textrm{(d)}}(h)e^{j\Delta \phi}}{r_{1,j}+r_{2,j}}\bigg|^{\alpha(h)}}}{{\sum \limits_{i \in\mathcal{B},  } 10^{\frac{G_e(\theta_{{\rm u},i})}{10}}\Bigg[ \frac{\sin^2\big({\frac{N_t\pi}{2}}\big(\sin(\theta_{{\rm u},i})-\sin(\phi_{{\rm u},i})\big)\big) } {\sin^2\big({\frac{\pi}{2}}\big(\sin(\theta_{{\rm u},i})-\sin(\psi_{{\rm u},i})\big)\big)}\big(\frac{1}{l_i} \big)\Bigg]}^{\alpha(h)} + \frac{1}{N_t}\sum \limits_{k \in\mathcal{B}, k \neq j}\bigg|10^{\frac{G_e(\theta_{{\rm d},k})}{10}}\frac{\sin^2\big({\frac{N_t\pi}{2}}\big(\sin(\theta_{{\rm d},k})-\sin(\phi_{{\rm d},k})\big)\big) } {\sin^2\big({\frac{\pi}{2}}\big(\sin(\theta_{{\rm d},k})-\sin(\phi_{{\rm d},k})\big)\big)} \big(\frac{1}{l_k} \big) + \frac{R(\psi_k) \widetilde{G}_k^{\textrm{(d)}}(h)e^{k\Delta \phi}}{r_{1,k}+r_{2,k}}\bigg|^{\alpha(h)}}$}.
\end{equation}
\hrulefill
\end{figure*}
Note that the rate of the UAVs associated with down-tilted antennas will be scaled by the parameter $\beta$. Lower values of $\beta$ will increase the SIR performance of the UAVs associated with the up-tilted antennas as shown in \eqref{eq:sir_uptilt_csf}. However, the UAVs associated with the down-tilted antennas and most importantly, the GUEs will suffer from low rates for a low $\beta$. This trade-off will be addressed in our future work.
\subsection{Problem definition} 

Our goal is to tune the UT angles of the up-tilted antennas individually during the USFs to provide reliable SIR at the UAVs' end. Without optimizing the UT angles, the SIR performance will worsen due to the additional interference from the up-tilted antennas~\cite{simran_GR}. Note that the UAVs can be associated with either up-tilted antennas or down-tilted antennas depending on the highest RSRP providing antenna set~\cite{simran_GR}. Let us consider the vector of SIRs of all UAVs when they are associated with the highest RSRP providing antenna sets as: 
 \begin{equation*}
      \mathbf{\gamma}=[ \gamma_{1,\rm usf},...,\gamma_{|\mathcal{A}|,\rm usf} ],
 \end{equation*}
 where $|\cdot|$ represents the cardinality of a set. Then, we can formulate the problem of maximizing the minimum UAV SIR as:
\begin{equation}
\label{eq:opt_prob}
\begin{aligned}
\max_{\mathbf{\Phi_{\rm u}}} \quad & \min \mathbf{\gamma}\\
\textrm{s.t.} \quad & 0 \leq  {\mathbf{\Phi_{\rm u}}} \leq 90^\circ.    \\
\end{aligned}
\end{equation}

Here, the optimization variable $\mathbf{\Phi_{\rm u}}=[\phi_{\rm u,1},...,\phi_{\rm u,{|\mathcal{B}|}}]$ is the vector of the UT angles of the up-tilted antennas in the network.
Note that only the interference caused by the up-tilted antennas is dependent on the UT angles. We also keep the UT angles above the horizon level (greater than $0^\circ$)~for saving the GUEs from additional interference. However, changing the UT angles will change the association of the serving GBS/antenna sets. Overall, the optimization problem in (18) is very difficult to solve efficiently since the objective function is highly non-convex with
respect to decision variables $\mathbf{\Phi_{\rm u}}$~\cite{geraci_2018}. The search space of the problem is continuous and grows exponentially with the number of GBS. Moreover, due to the complex antenna pattern and tilting angles involved, it is not possible to obtain the closed-form optimal solutions by taking the derivatives of \eqref{eq:sir_uptilt_closed} and \eqref{eq:sir_downtilt_closed} even under a free-space path-loss model and a similar UT angle for all the GBSs. Assuming the tilting angles to be $0^\circ$ for simplification as done in~\cite{ramy_saad} will not represent a realistic cellular network scenario. 
\vspace{-0.1cm}
Using an exhaustive search method is also computationally prohibitive since its complexity increases exponentially with number of GBSs or up-tilted antenna sets. To overcome these challenges, in the next section, we introduce our GA-based UT angle optimization method for maximizing the minimum UAV SIR. Note the SIR gain due to the eICIC is not related to tuning the UT angles and the gain can be calculated by simply not considering the received power from the down-tilted antennas. The rates of the UAVs who are associated with the down-tilted antennas will be reduced by the quantity $\beta$ as shown in \eqref{eqn:rate_downtilt} and their SIRs will also be impacted by the choice of the UT angles.

\section{Genetic Algorithm-Based Up-tilt Angle Optimization}
\label{sec:GA}
The GA is a stochastic population-based optimization technique that mimics the metaphor of natural biological evaluation and is an efficient tool in searching for the global optimum~\cite{GA_book}. It borrows the idea of “survival of the fittest” in its search process to select and generate individuals (design solutions) that are adapted to the underlying objectives/constraints of the problem of interest. Hence, GA is well suited to and has been extensively applied to solve complex design optimization without being guided by stringent mathematical formulation. It can explore the whole search space simultaneously, and hence, identify high quality solutions more quickly than an exhaustive search. The detailed principles of a GA scheme can be found in~\cite{GA_book}. In the following subsections, we outline our proposed GA-based UT angle tuning method for obtaining the optimal solution of \eqref{eq:opt_prob}. We assume that each GBS sends only its chosen UT angle and the SIR information of the UAVs associated with it to a central server. The server can then run the proposed GA-based algorithm and compute the optimum UT angles.

\subsection{Representation}
At first, some randomly generated candidate solutions for the optimization problem are encoded in a chromosome-like strings. The collection of these candidate solutions or chromosomes are referred to as population. In other words, members of the population are the vectors of possible UT angles for our formulated optimization problem. Note that each member of the population must provide a complete solution to the problem. The size of the population does not change over time usually. To meet the constraint, the UT angles of the population are generated within the feasible search space.

\subsection{Fitness evaluation}
The objective function of the problem is used to evaluate the fitness of each chromosome. In our case, the randomly generated UT angles are used as inputs to the simulator for obtaining the minimum SIR of all the discrete UAV locations. The higher the minimum SIR of a solution is, the better the fitness value is associated with it.

\subsection{Selection}
The selection process determines the pair of candidate solutions/ UT angles which will act as parents for mating. After being evaluated by a \emph{fitness function}, each member of the population is assigned a probability to be selected for reproduction. Note that, the worse performing members should also be given a chance in the evolution process so that the overall algorithm can maintain a good exploration in the search space. Here, we consider a simple biased roulette wheel to select individuals as parents~\cite{elhachmi2016cognitive}. More explicitly,  each chromosome in the population is assigned a slot in a roulette wheel, whose size is proportional to its fitness over the total sum of fitness in the population. Then, a random number between $0$ and $1$ is generated for each member/ UT angle set. A chromosome/member is selected as a parent for further genetic operations if the random number is within the range of its roulette wheel slot. 

\begin{algorithm}[!t]\small 
	\caption{Up-tilt Angle Optimization using GA}
    \label{alg:Alg_GA}
	\begin{algorithmic}[1]
		\STATE \textbf{Input:} 
		\STATE population: Set of UT angles for all GBSs
		\STATE   Fitness function (FF): Minimum SIR of the UAV
		\STATE   network parameters, GBS and UAV locations 
		\STATE \textbf{Method:}
		\STATE NewPopulation $=$ empty set
		\STATE StopCondition: Number of iterations
		\STATE SELECTION: Roulette wheel selection method
		\STATE Create random Population
		\STATE EVALUATE~(Population, FF)
		\STATE  \textbf{while}~(StopCondition is not met)
		\STATE  \hspace{0.3cm} \textbf{for}  $i=1 \;\text{to}\;$ Population size \textbf{do}
		
		\STATE  \hspace{0.6cm} Parent1 = SELECTION(NewPopulation, FF)
		\STATE  \hspace{0.6cm} Parent2 = SELECTION(NewPopulation, FF)
		\STATE  \hspace{0.6cm} Child = Reproduce(Parent1, Parent2)
		\STATE  \hspace{0.6cm} \textbf{if} (small random probability)
		\STATE  \hspace{0.6cm} child = MUTATE(Child)
		\STATE  \hspace{0.6cm} add child to NewPopulation set
        \STATE  \hspace{0.6cm} \textbf{end if}
        \STATE  \hspace{0.3cm} \textbf{end for}
		\STATE  \textbf{end while}
		\STATE EVALUATE~(NewPopulation, FF)
		\STATE Args = GetBestSolution~(NewPopulation)
		\STATE Population = Replace~(Population, NewPopulation)
		\STATE \textbf{Output:} Args: Best individuals of the UT angles and the highest minimum SIR
		
	\end{algorithmic}
\end{algorithm}

\subsection{Crossover}
The selected parents are then processed by the crossover operator, which mimics mating in biological populations. It is considered to be the most significant phase in a GA. Here, for each pair of parents to be mated, a crossover point is chosen at random from within the chromosomes. Then offspring/children are created by exchanging the chromosomes (UT angles) of parents among themselves until the crossover point is reached. The crossover operator propagates features of good surviving designs from the current population into the future population, which will have better fitness value (higher minimum SIR in our case) on average.

\subsection{Mutation}
The last operator is the mutation, which introduces diversity in population characteristics and prevents premature convergence. In this step, certain parts of the newly formed children (new sets of UT angles with better fitness) are subjected to a mutation with a low random probability. In our proposed GA-based framework, the mutation takes place with a low mutation probability. We first generate random numbers between $-1$ and $1$ for each member of the UT angle population. If the absolute value of a random number is less than the mutation probability, that particular random number is added to that member (UT angle) of the population. 

After all of these genetic processes, the members of the populations with the worst fitness values are replaced by the new individuals with better fitness values or higher minimum SIRs. The algorithm continues until good results are obtained through iterations in terms of the objective function. The overall algorithm is also summarized in Algorithm~\ref{alg:Alg_GA}. In essence, obtaining high-quality suboptimal solutions from our proposed method depends on carefully addressing the following issues.
\begin{itemize}
    \item representation of tentative solutions (UT angles) as chromosomes;
    \item initialization of the randomly generated population;
    \item determination of the fitness function (min SIR);
    \item selection of genetic operators;
    \item adjustment of GA parameters (population size, crossover
and mutation probabilities).
\end{itemize} 

Considering the impact of mutation, the work in~\cite{greenhalgh2000convergence} provided the lower bound of the number of iterations required for obtaining the global optimum for a given population size. In particular, they showed that to obtain the global optimum with any specified level of confidence, GAs should run for long enough. However, later we show that increasing the number of iterations or population size will increase the complexity and run-time of the proposed algorithm. Hence, we run extensive simulations for different numbers of population size and iterations, and check the associated minimum UAV SIRs. We found that with a the population size of $200$, mutation probability of $0.1$, and $50$ iterations, our algorithm provides high-quality suboptimal solutions.

\subsection{Complexity analysis}

As described in the previous subsections, our proposed GA-based UT angle optimization technique randomly generates tentative solutions and then produces new better solutions from the previous ones iteratively. For a given GBS and UAV distributions, the overall time complexity of the algorithm is $\mathcal{O}(M^2I|\mathcal{A}||\mathcal{B}|)$, where $M$ represents the number of populations and $I$ is the iteration number, respectively. Hence, for a given population size, number of iterations, and number of GBSs, the complexity of our proposed algorithm increases linearly with an increasing number of UAVs. 

\begin{figure}[!t]
\centering
		\subfloat[$h_{\textrm{UAV}}=100$~m, ISD~$=500$~m.]{
			\includegraphics[width=\linewidth]{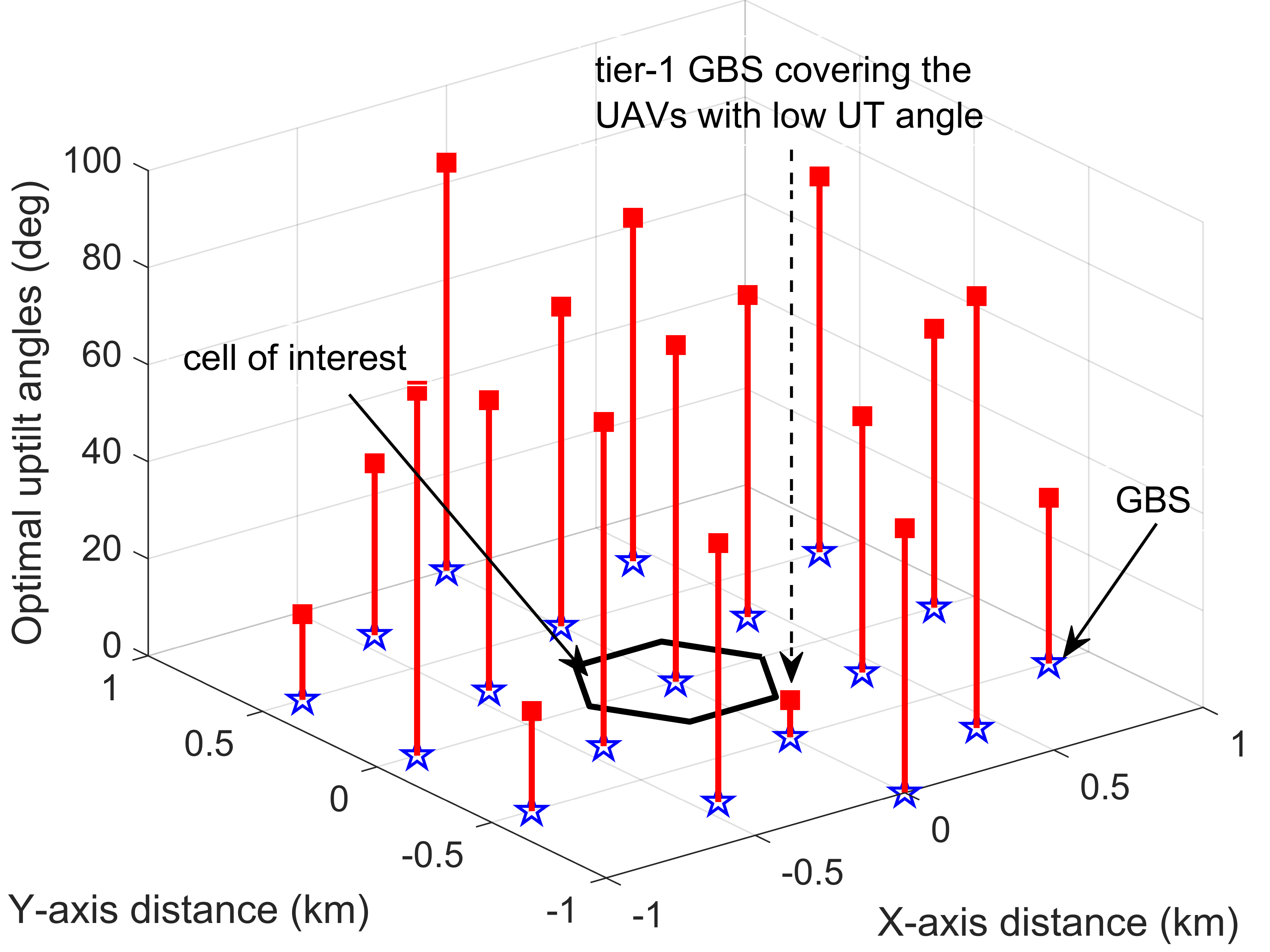}} 
	\\	\subfloat[$h_{\textrm{UAV}}=200$~m, ISD~$=500$~m.]{
			\includegraphics[width=\linewidth]{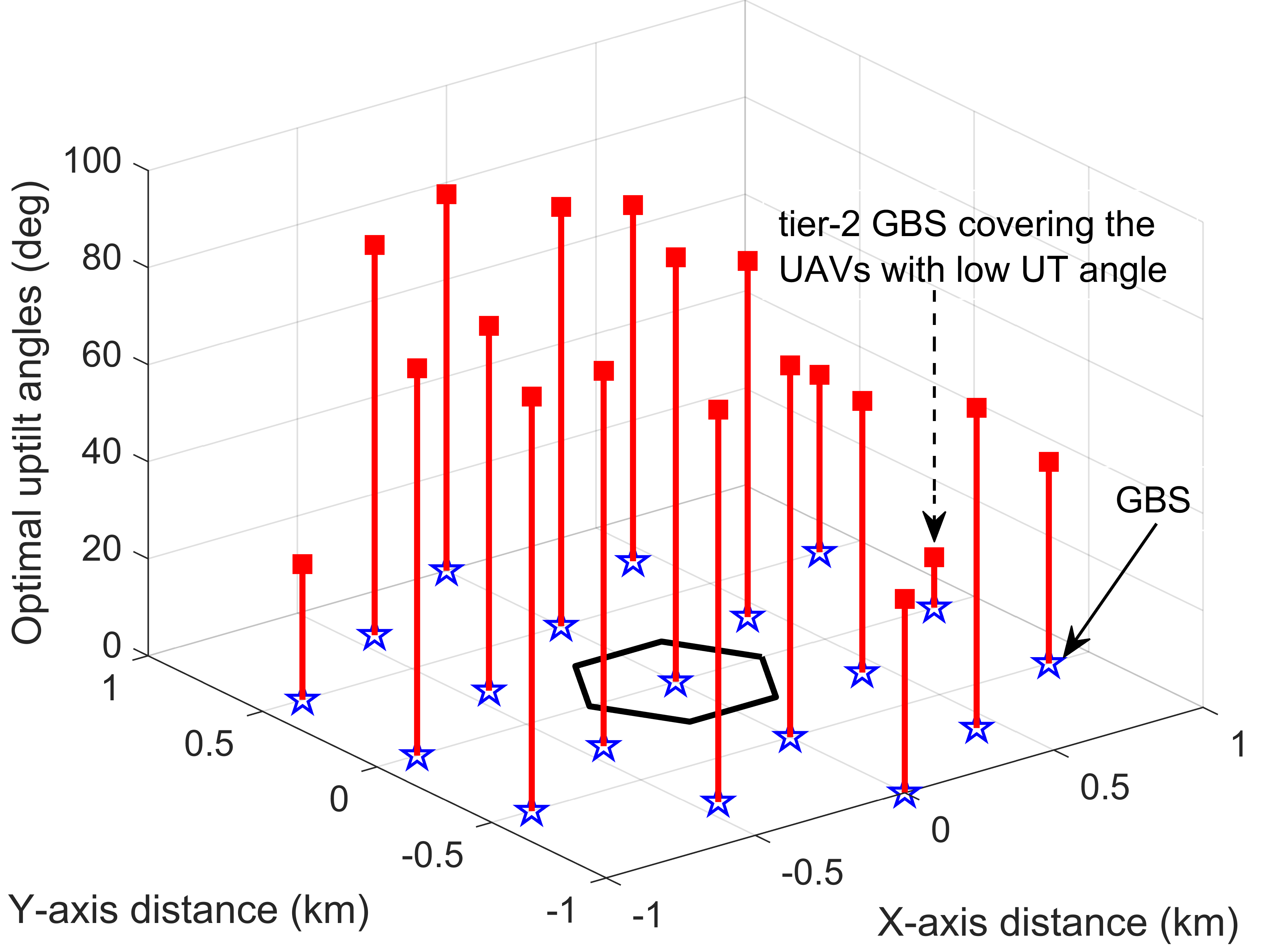}} 
	\\	\subfloat[$h_{\textrm{UAV}}=100$~m, ISD~$=1000$~m.]{
		\includegraphics[width=\linewidth]{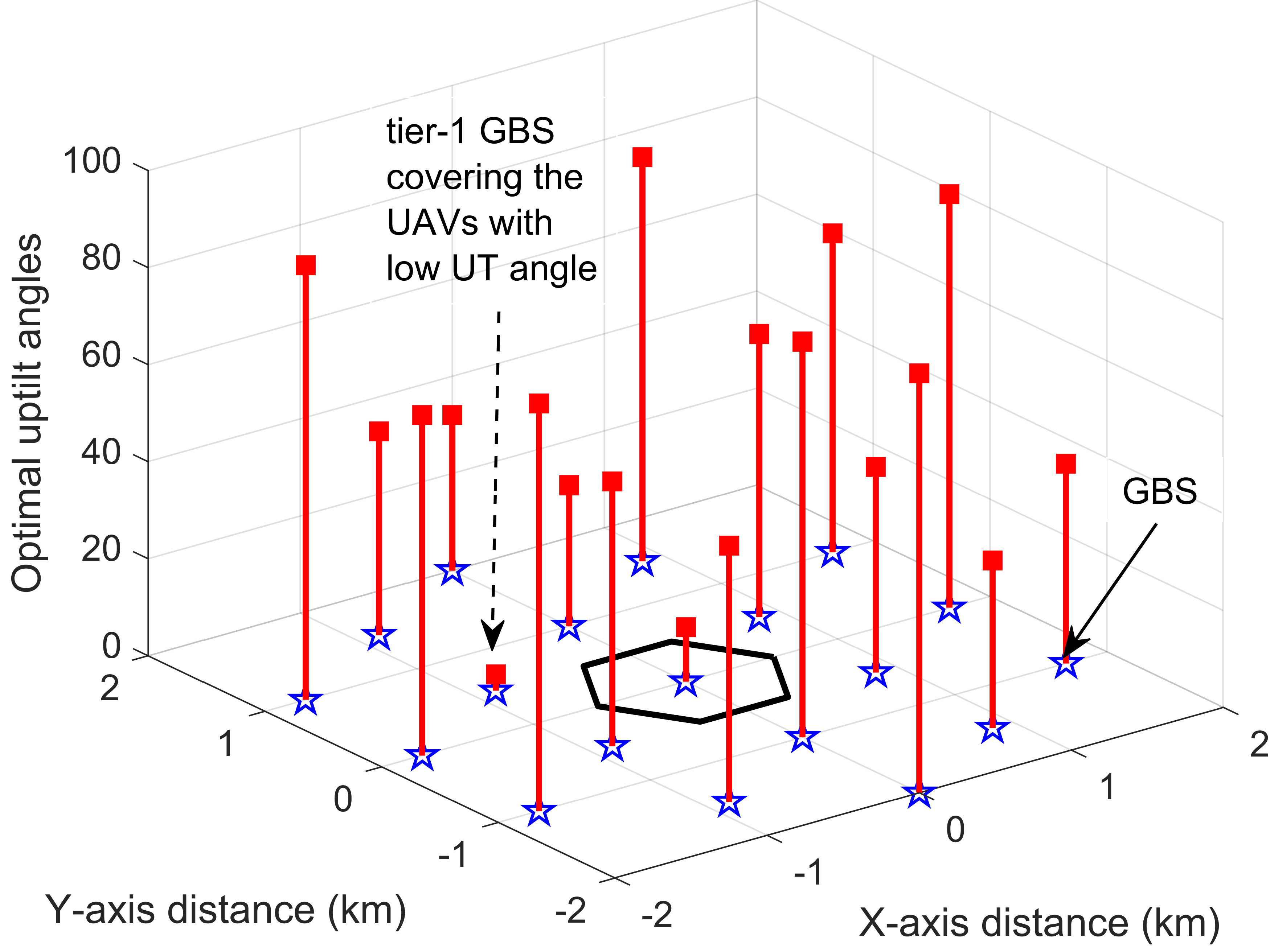}}
		\caption {
		{Optimal UT angles obtained from the proposed GA algorithm for ISD~$=500$~m for (a) $h_{\textrm{UAV}}=100$~m, (b) $h_{\textrm{UAV}}=200$~m, and (c) for ISD~$=1000$~m and $h_{\textrm{UAV}}=100$~m.}}
		\label{fig:opt_angles_500mISD}
\end{figure}

\section{Simulation Results}
\label{sec:simu_rl}

\begin{table}[t]
\centering
\renewcommand{\arraystretch}{1}
\caption {Simulation parameters.}
\scalebox{1}
{\begin{tabular}{lc}
\hline
Parameter & Value \\
\hline
$P\textsubscript{GBS}$ & 46 dBm  \\ 
$h\textsubscript{UAV}$ & $100$~m \& $200$~m\\ 
$h_{\rm GBS}^{\rm (d)}$ & $30$ m\\ 
ISD & $500$~m \& $1000$~m\\ 
$h_{\rm d}$ & $1$ m\\ 
$h\textsubscript{GUE}$ & $1.5$ m\\ 
$\lambda$ & $0.15$ m\\ 
$\alpha_0$ & $3.5$~\cite{wietfeld_ground_reflection}\\ 
DT angle ($\phi_{\rm d}$) & $6^\circ$ \\
\hline
\end{tabular}}
\label{tab:sim_para}
\end{table}
In this section, we present the simulation results for our proposed cellular architecture based on a new set of antennas and eICIC. Unless otherwise stated, the simulation parameters are as listed in Table~\ref{tab:sim_para}. By considering flat fading channels~\cite{guvenc_letter_eICIC} and hexagonal cells, we report our finding for two ISDs namely, $500$~m and $1000$~m while considering the highest RSRP-based association (HRA). It is worth noting that in our setup, the HRA association will also provide the highest SIR among all the available antennas of the network. For convenience, we refer to our proposed method as `optimal HRA' hereinafter. To study the performance of our proposed method we consider also three baseline schemes. These four scenarios can be summarized as follows.
\begin{itemize}
    \item \textit{optimal HRA}: this is our proposed GA-based UT angle tuning method.
    \item \textit{HRA single}: all GBSs pick the same optimal UT angle which maximizes the minimum SIR. This UT angle is calculated by exhaustive search method.
    \item \textit{Random}: each GBS picks UT angles randomly from the search space.
    \item \textit{HRA (no eICIC nor UT antennas)}: presence of up-tilted antennas and eICIC is ignored. UAVs associate with the highest RSRP providing GBS.
\end{itemize}

As mentioned in Section~\ref{sec:sys}, we divide the whole network into $10$~m$\times10$~m grids~\cite{guvenc_letter_eICIC}, and a UAV is placed on each grid point with height $h_{\rm UAV}$. Such a uniform distribution will average out the impact of UAV distributions~\cite{guvenc_letter_eICIC}. We only take the discrete points inside the center hexagonal cell into consideration. 

\subsection{Optimal UT angle analysis}
After obtaining the best solutions of UT angles by using \eqref{eq:sir_uptilt} and \eqref{eq:sir_downtilt} and our proposed GA-based method, we calculate the UAV SIRs in USFs for the two ISDs and UAV heights. Then eICIC is used to get the pertinent UAV SIRs in CSFs. For ISD~$=500$~m and $h_{\textrm{UAV}}=100$~m and $200$~m, the best solutions obtained from the proposed GA-based algorithm are presented in Fig.~\ref{fig:opt_angles_500mISD}(a) and Fig.~\ref{fig:opt_angles_500mISD}(b), respectively. Our results show that one of the six neighboring GBS chooses a relatively smaller UT angle and provides high received power to the UAVs for $h_{\textrm{UAV}}=100$~m. The other GBSs overall maintain higher UT angles to reduce the interference from the side lobes. 

A similar conclusion can also be drawn for $h_{\textrm{UAV}}=200$~m, while one big exception is that the UAVs are supported by s tier-2 GBS as shown in Fig.~\ref{fig:opt_angles_500mISD}(b). Due to the compact GBS locations and higher UAV height, the tier-2 GBSs can provide better SIR by choosing an angle that covers most of the discrete UAV locations for $h_{\textrm{UAV}}=200$~m. For ISD $=1000$~m, both UAV heights show the similar trend as Fig.~\ref{fig:opt_angles_500mISD}(a) and in Fig.~\ref{fig:opt_angles_500mISD}(c), we report the best solutions of UT angles for $h_{\textrm{UAV}}=100$~m. Overall, the GBSs tend to choose lower UT angles for larger ISD to reduce inter-cell interference. A similar case of obtaining lower UT angles for higher ISD was also reported in~\cite{lin2020a2g}.

\begin{figure}[t]
\centering
		\subfloat[$h_{\textrm{UAV}}=100$~m.]{
			\includegraphics[width=\linewidth]{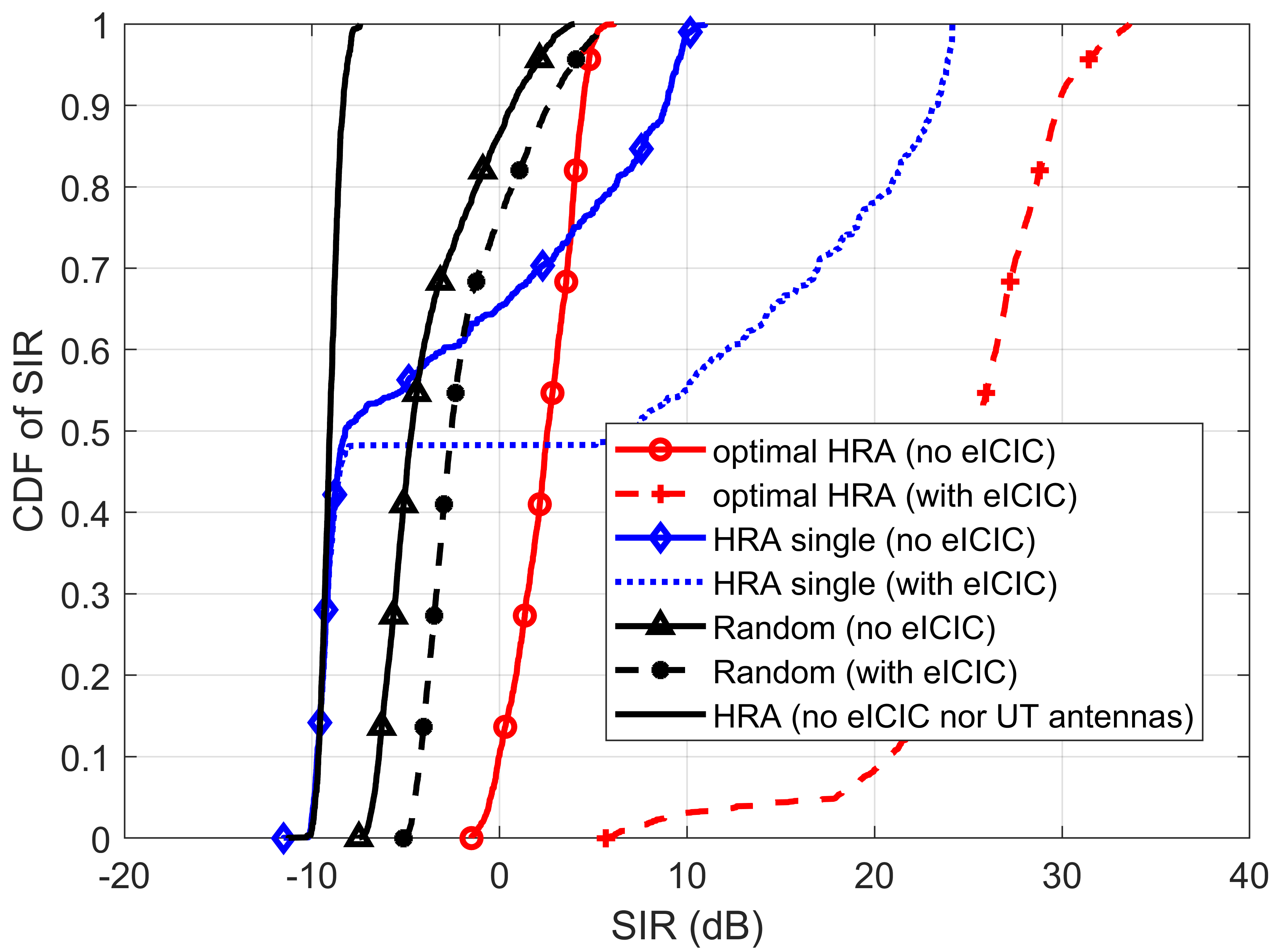}} 
			\hfill
		\subfloat[$h_{\textrm{UAV}}=200$~m.]{
			\includegraphics[width=\linewidth]{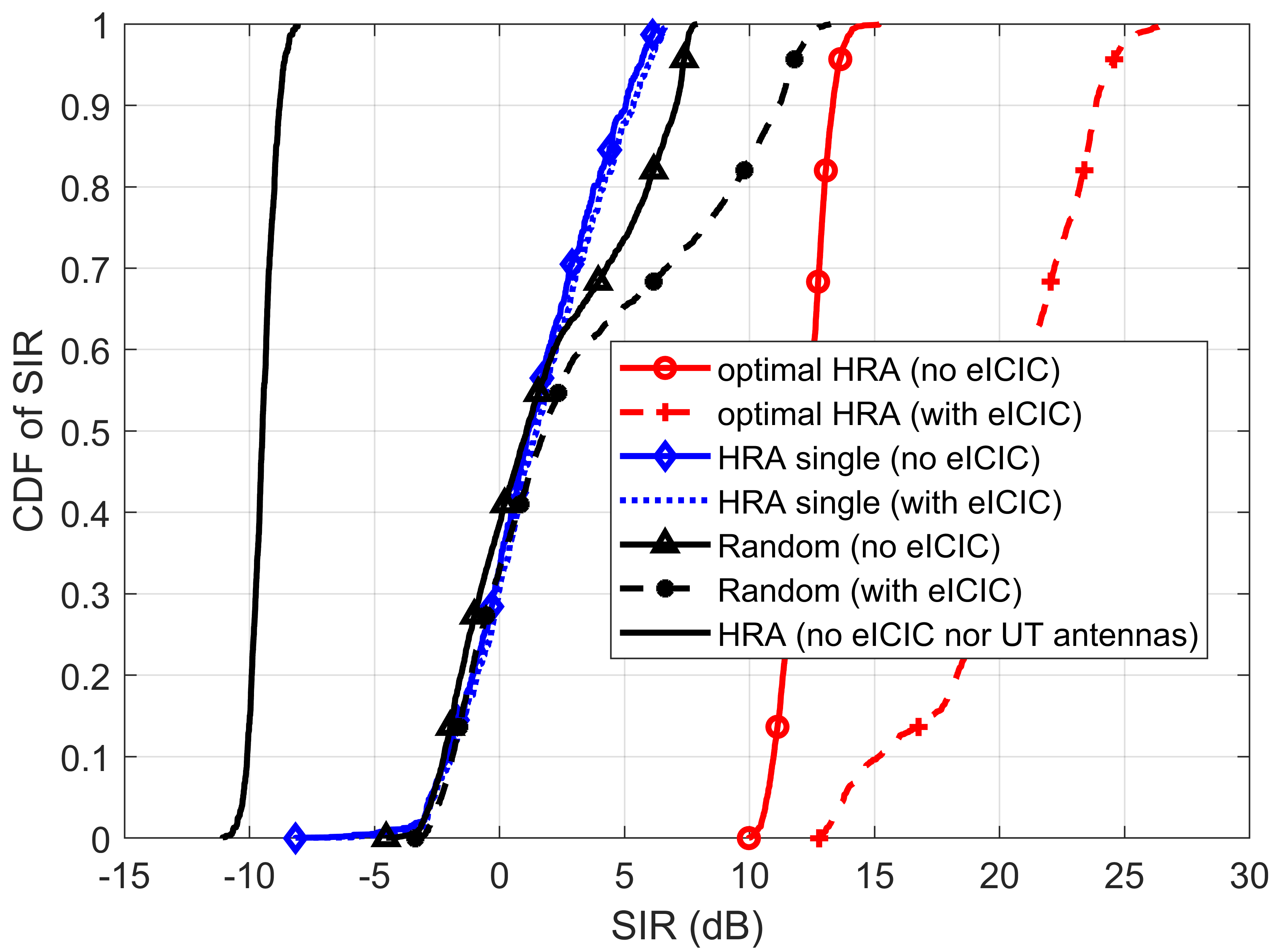}}
		\caption {
		{UAV SIR CDFs for  ISD~$=500$~m for (a) $h_{\textrm{UAV}}=100$~m and (b) $h_{\textrm{UAV}}=200$~m.}}
		\label{fig:sir_500mISD}
\end{figure}

For ISD $=500$~m and $h_{\textrm{UAV}}=100$~m and $200$~m, the respective UAV SIR cumulative distribution function (CDF) plots are presented in Fig.~\ref{fig:sir_500mISD}(a) and Fig.~\ref{fig:sir_500mISD}(b), respectively. From both figures, we can conclude that our proposed optimal HRA scheme provides higher minimum SIR (about $-1.36$~dB for $h_{\textrm{UAV}}=100$~m and about $10$~dB for $h_{\textrm{UAV}}=200$~m) than the other baseline methods. The optimization framework considers the minimum UAV SIR inside the center cell and thus the interfering GBSs choose UT angles which create less interference towards the UAVs. During the CSFs, turning the down-tilted antennas off increases the minimum SIR to about $6$~dB for $h_{\textrm{UAV}}=100$~m and about $12.5$~dB for $h_{\textrm{UAV}}=200$~m. One interesting observation is that the overall SIR with eICIC is higher for $h_{\textrm{UAV}}=100$~m. This is because the UAVs suffer more interference from the down-tilted antennas for lower UAV heights via GR and antenna side lobes. Moreover, the path-loss is also lower for $h_{\textrm{UAV}}=100$~m than $h_{\textrm{UAV}}=200$~m. Hence, muting the down-tilted antennas provide higher SIR gain in the CSFs for $h_{\textrm{UAV}}=100$~m.

\begin{figure}[!t]
\centering
		\subfloat[$h_{\textrm{UAV}}=100$~m.]{
			\includegraphics[width=\linewidth]{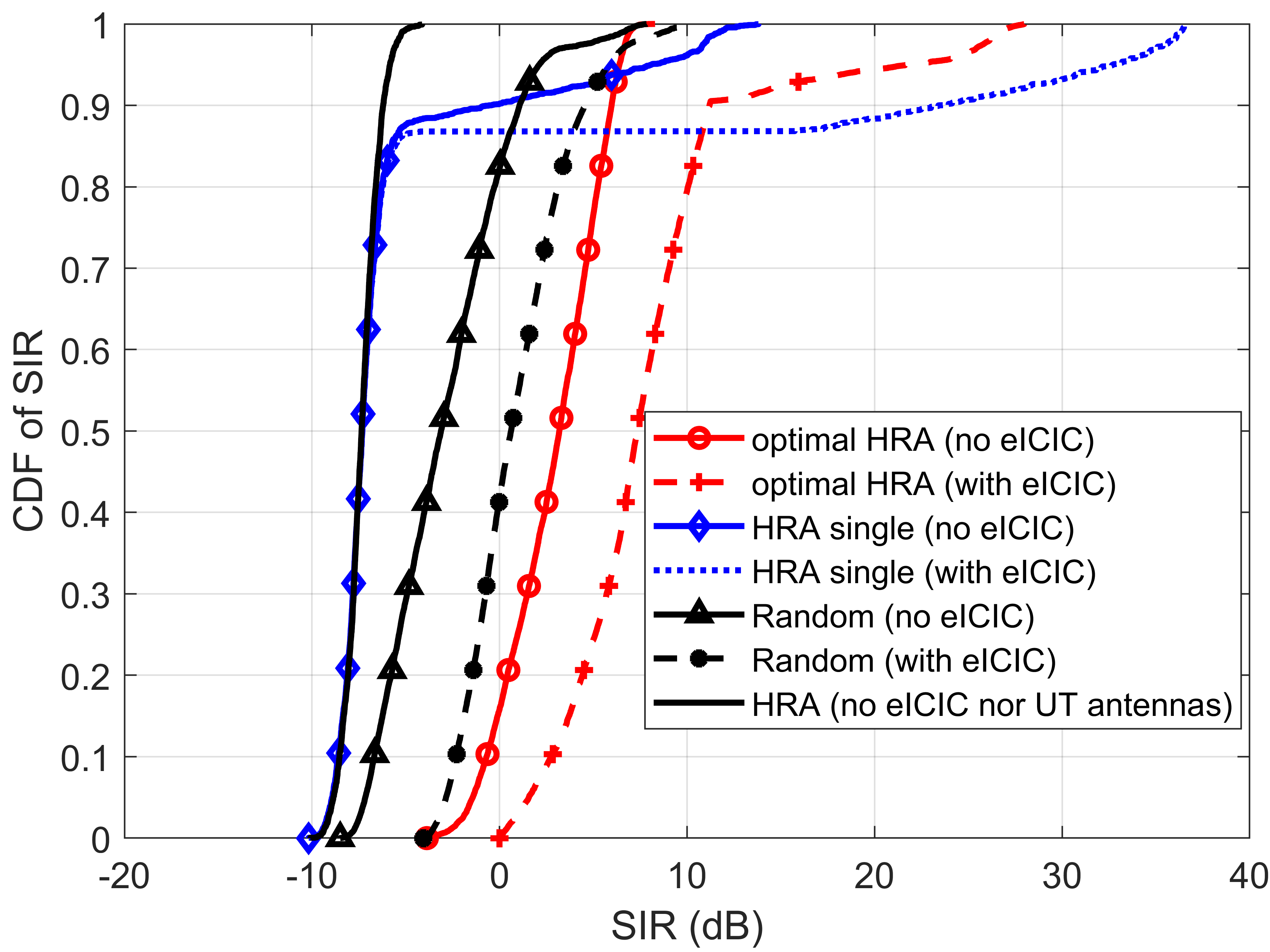}} \hfill
		\subfloat[$h_{\textrm{UAV}}=200$~m.]{
			\includegraphics[width=\linewidth]{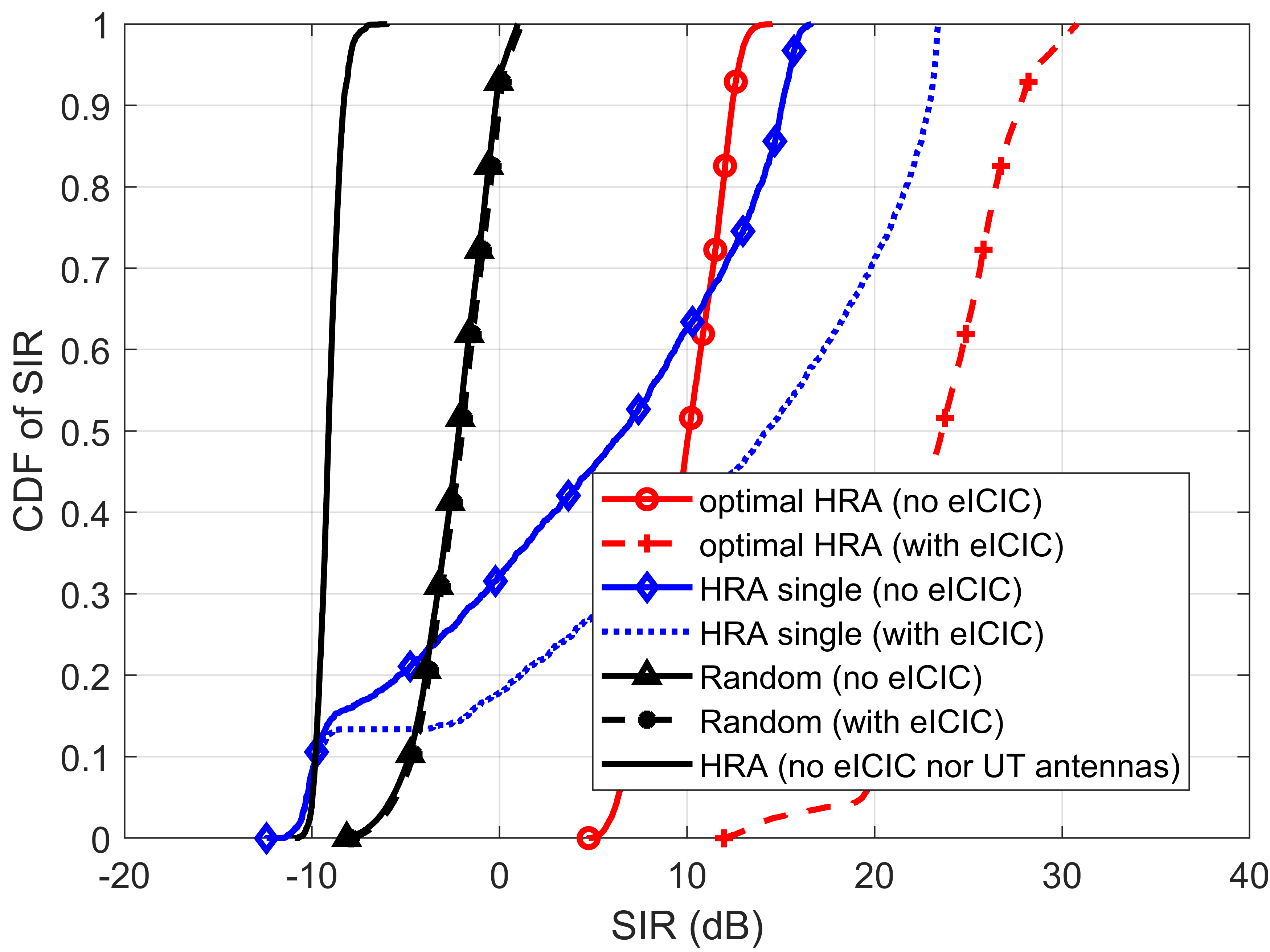}}
		\caption {
		{UAV SIR CDFs for $D=1000$~m for (a) $h_{\textrm{UAV}}=100$~m and (b) $h_{\textrm{UAV}}=200$~m.  }}
		\label{fig:sir_1000mISD}
\end{figure}

\begin{figure}[!t]
\centering
		\subfloat[]{
			\includegraphics[width=\linewidth]{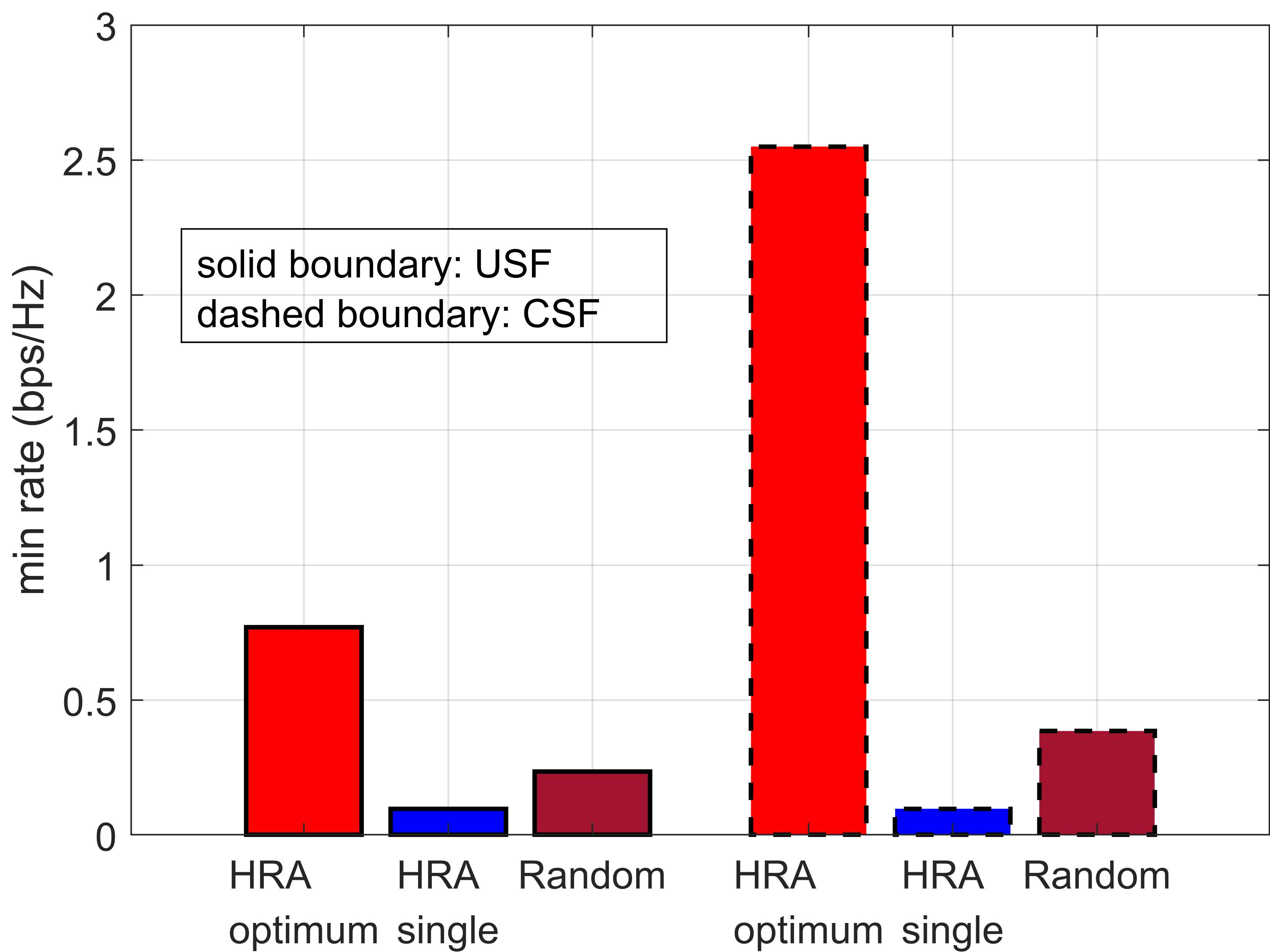}} 
	\\	\subfloat[]{
			\includegraphics[width=\linewidth]{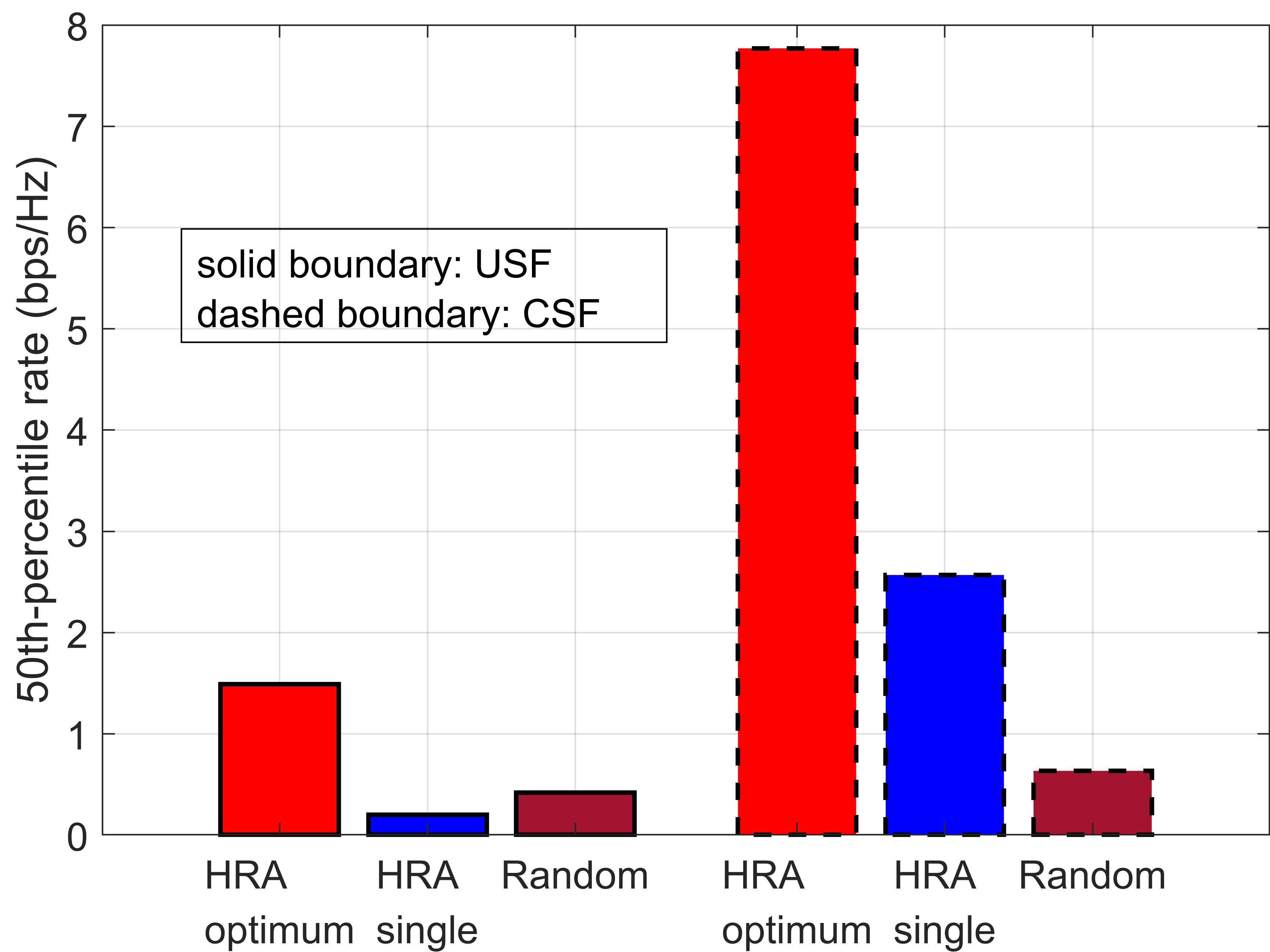}} 
	\\	\subfloat[]{
		\includegraphics[width=\linewidth]{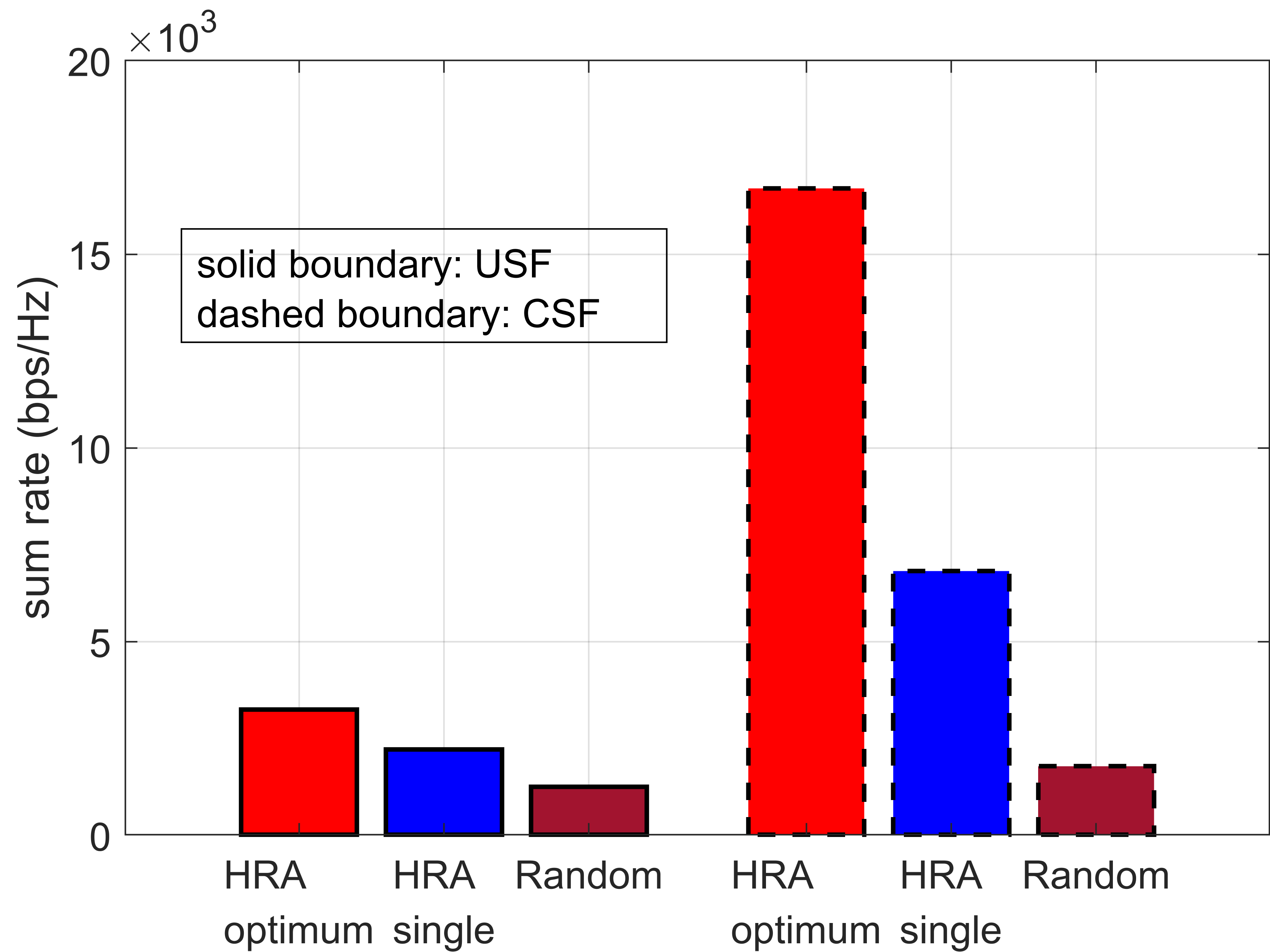}}
		\caption {
		{Rate (bps/Hz) analysis for $h_{\textrm{UAV}}=100$~m and ISD~$=500$~m. (a) min rate, (b) $50$th-percentile rate, and (c) sum rate.}}
		\label{fig:rate_analysis}
\end{figure}

In the HRA single scheme, the GBSs choose the same optimal angle, which result in less degree of freedom to improve the SIR performance. Hence, it provides comparatively lower SIR (about $-11$~dB for $h_{\textrm{UAV}}=100$~m and about $-8$~dB for $h_{\textrm{UAV}}=200$~m) than our proposed method. Even with the ICIC, the overall gain in the minimum SIR is still significantly lower than without the ICIC minimum SIR of our proposed scheme. The random scheme chooses the UT angles for each of the GBSs and thus provides better performance than HRA single. Thus, \emph{it is evident from the discussion that it is critical to tune the UT angles of the GBSs individually for the successful integration of the up-tilted antenna sets}. Finally, for the case in which the UAVs are served by only down-tilted antennas and without the ICIC scheme, the overall SIR is very low (less than $-8$~dB) for both of the UAV heights. For larger cell sizes or ISD $=1000$~m and the two UAV heights, we can conclude from Fig.~\ref{fig:sir_1000mISD}(a) and Fig.~\ref{fig:sir_1000mISD}(b) that our method outperforms the other baseline schemes significantly in terms of the minimum UAV SIR during the USFs i.e., without ICIC. 
Fig.~\ref{fig:rate_analysis} shows the rates (bps/Hz) for the baseline schemes using \eqref{eqn:rate_uptilt} and \eqref{eqn:rate_downtilt}. From Fig.~\ref{fig:rate_analysis}, we can observe that our proposed optimal HRA scheme provides a higher minimum rate, $50$th-percentile rate, and sum rate than other baseline schemes. The HRA (no ICIC or UT antennas) scheme is excluded in the rate comparison due to its very low SIR performance (less than $-8$~dB). Due to the higher SIR obtained with eICIC, overall the rates increase significantly in the CSFs. The UAV with the minimum SIR in the HRA single scheme is associated with the down-tilted antennas and thus, HRA single provides the same rate in USF and CSF. Similar observations are also obtained for other UAV height and ISD. 

\subsection{Impact of the down-tilted antenna}
\begin{figure}[!t]
\centering
		\subfloat[$h_{\textrm{UAV}}=100$~m.]{
			\includegraphics[width=\linewidth]{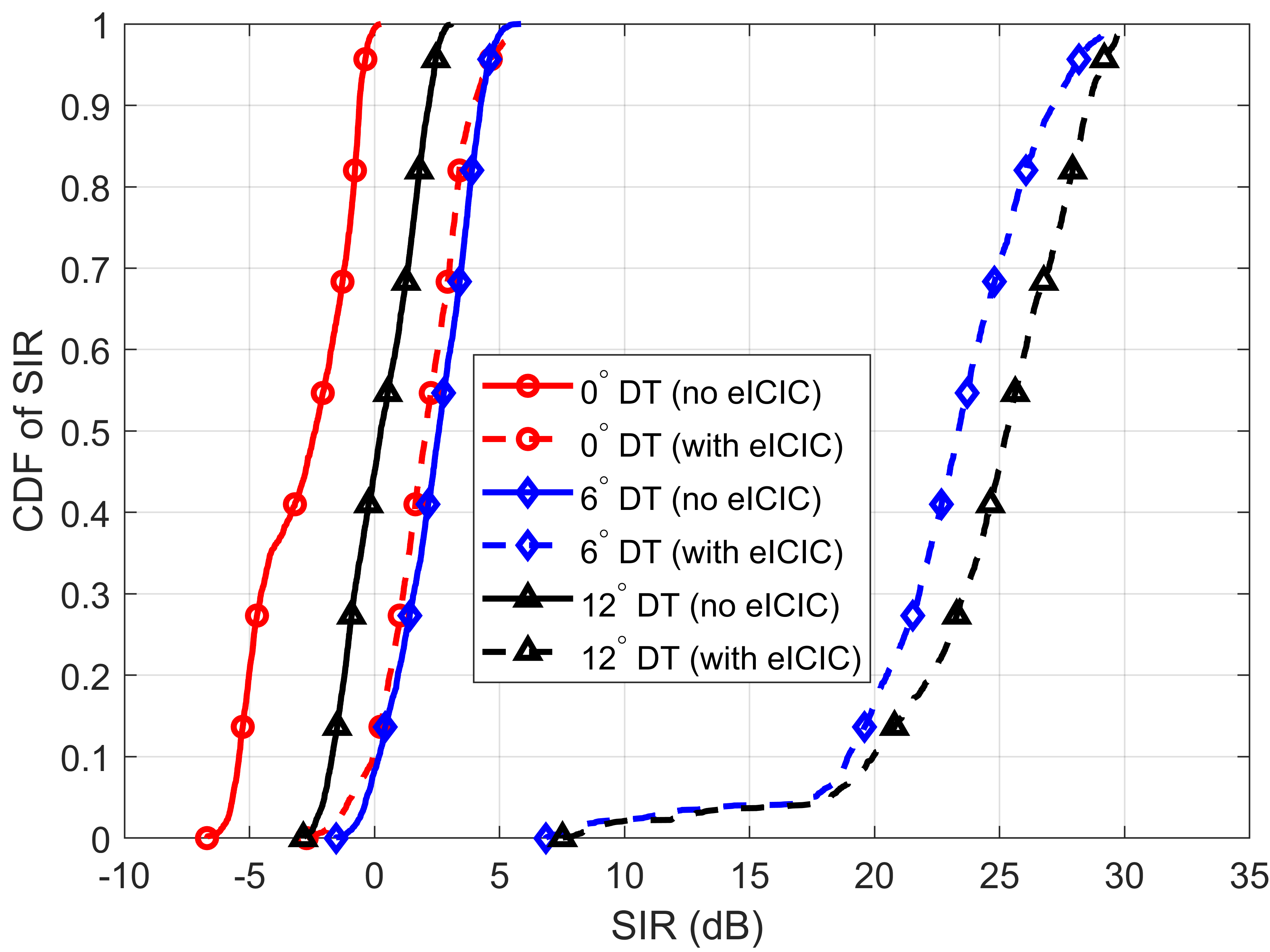}} \hfill
		\subfloat[$h_{\textrm{UAV}}=200$~m.]{
			\includegraphics[width=\linewidth]{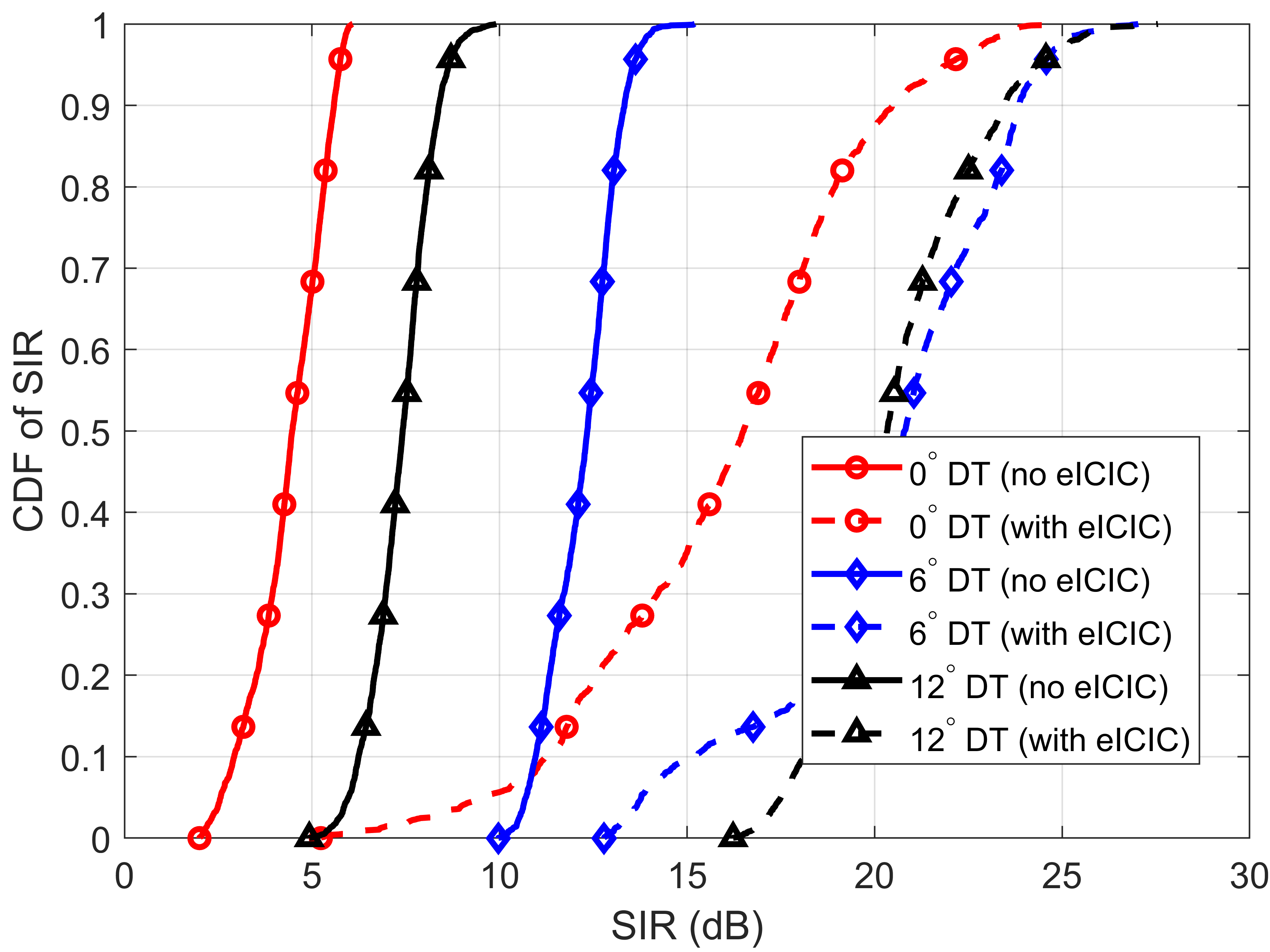}}
		\caption {
		{UAV SIR CDFs for ISD~$=500$~m for (a) $h_{\textrm{UAV}}=100$~m and (b) $h_{\textrm{UAV}}=200$~m. }}
		\label{fig:impact_DT}
\end{figure}

DT angles can create a significant impact on the overall performance of the network since they play a major role in determining the inter-cell interference. Higher DT angles decrease the interference towards other nearby GBSs which translates to a better coverage for GUEs. However, for UAVs flying in the sky, the DT angles can create interference by both side lobes and GR. This motivated us to study the impact of DT angles of the down-tilted antenna sets and report the relevant results in Fig.~\ref{fig:impact_DT}.

In Fig.~\ref{fig:impact_DT}(a), we show the SIR CDFs for $h_{\rm UAV}=100$~m and $200$~m by calculating the optimal UT angles using an optimal HRA scheme for three DT angles namely, $0^\circ$, $6^\circ$, and $12^\circ$, respectively. From this figure, we can conclude that the $0^\circ$~DT angle overall provides low SIR in both USF and CSF frames due to the higher interference stemming from the main beam of the down-tilted antennas. Though the impact of GR is trivial for $\phi_{\rm d}=0^\circ$ as discussed in Theorem 1, the focus of the main beam causes severe interference to the far away UAVs, which degrades the overall SIR performance. Although higher DT angles are beneficial for GUEs, our results show that $6^\circ$ provides better SIR performance than its $12^\circ$ counterpart. This is because, for a $12^\circ$ DT angle, the UAVs faces more interference by GR from the closest GBS as described in Theorem 1. For a $6^\circ$ DT angle, UAVs usually suffer less severe interference in GR from neighbor GBSs due to higher path-loss since the GR signals have to travel longer to reach the UAV. 

For the CSFs, we obtain high SIR for both $6^\circ$ and $12^\circ$. Due to the higher GR interference of $12^\circ$, this angle provides the highest SIRs in the CSFs by muting the down-tilted antennas. From Fig.~\ref{fig:impact_DT}(b), we can make similar observations for $h_{\rm UAV}=200$~m. However, in Fig.~\ref{fig:impact_DT}(b), the UAVs achieve better SIRs than those of lower heights. This is due to the fact that the GRs from the GBSs face higher path-loss and thus become weak when they reach UAVs. Moreover, the interference due to the side lobes also weakens due to the increased distances from the GBSs. Interestingly, $6^\circ$ provides slightly better SIRs because this angle provides better antenna gain through the side lobes from its other DT angle counterparts at $h_{\rm UAV}=200$~m. 

\subsection{Impact of the number of antenna elements}

\begin{figure}[!t]
\centering
		\subfloat[]{
			\includegraphics[width=\linewidth]{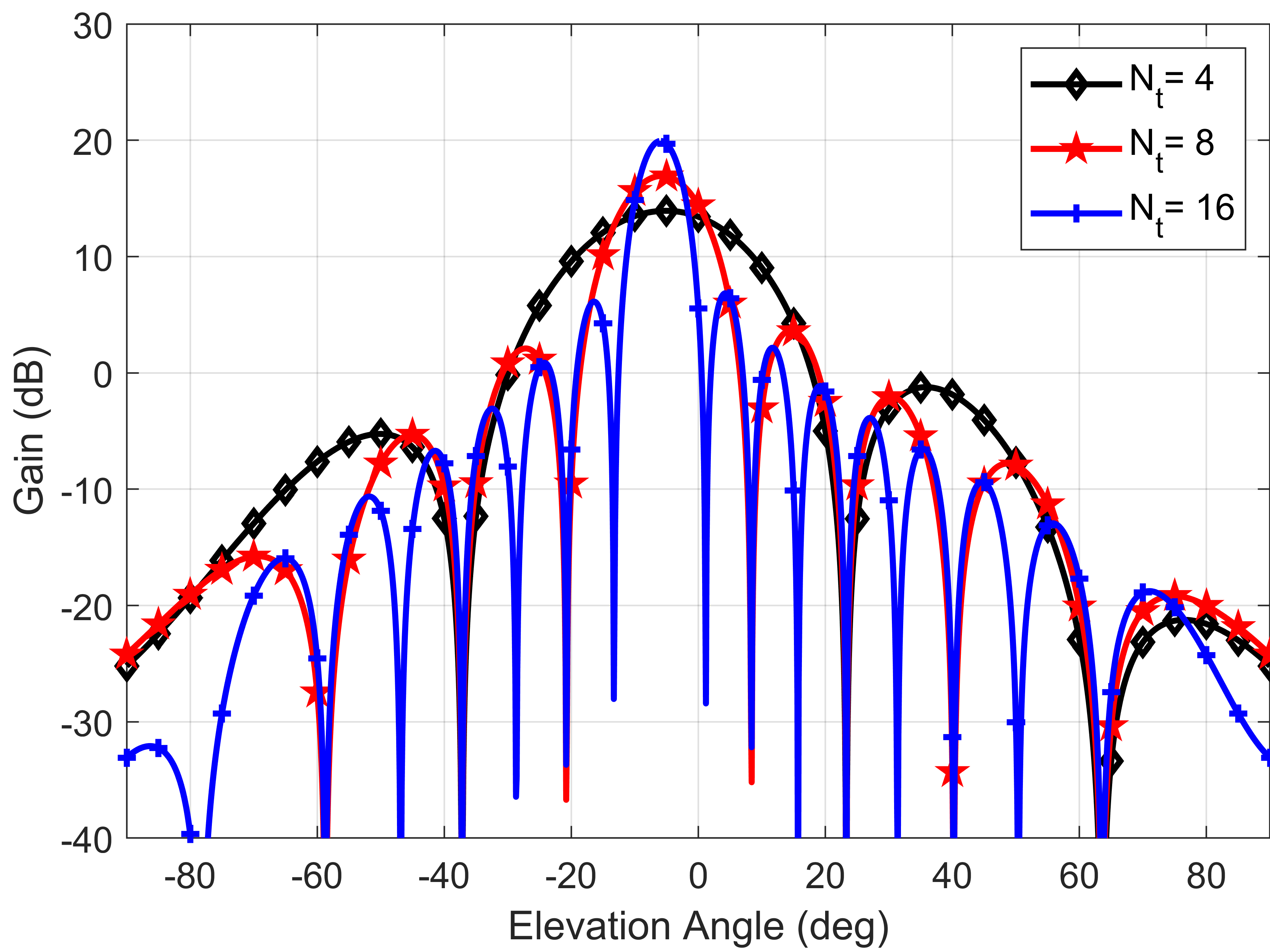}} \hfill
		\subfloat[]{
			\includegraphics[width=\linewidth]{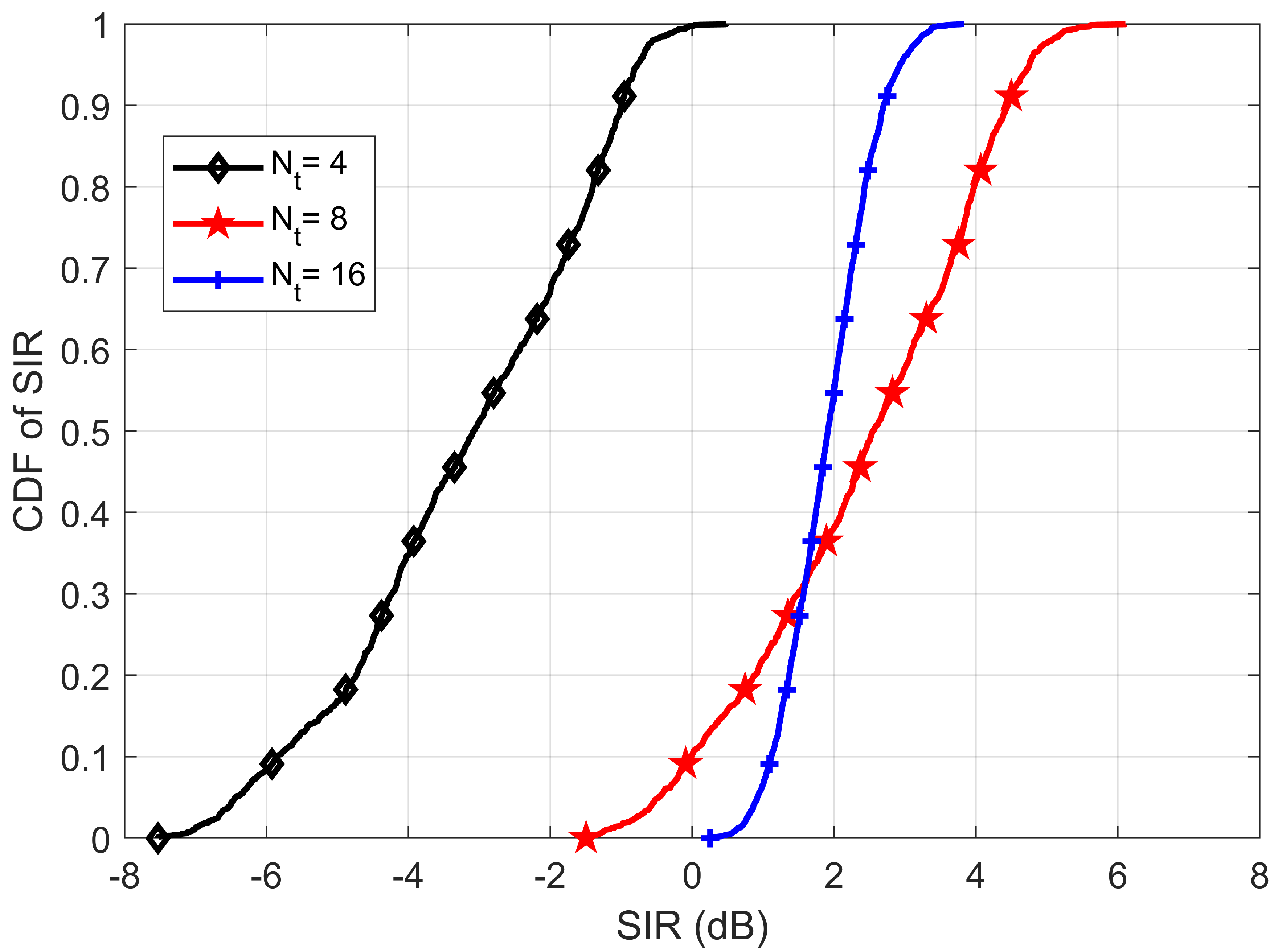}}
		\caption {
		{(a) Vertical antenna pattern of a GBS considering cross-polarized elements, each
with $65^\circ$ half power beam width and $\phi_{\rm d}=6^\circ$. (b) UAV SIR CDFs for $h_{\textrm{UAV}}=100$~m and  ISD~$=500$~m during the USFs. }}
		\label{fig:impact_element}
\end{figure}

The number of antenna elements has a direct impact on the antenna array gain and the beam width of the antenna pattern~\cite{antenna_gain_beamwidth}. Here, we focus on how the number of antenna elements at the GBS can influence the SIR performance of the UAVs. Note that increasing the element number increases the antenna array gain but reduces the beam width and vice versa~\cite{antenna_gain_beamwidth}. In Fig.~\ref{fig:impact_element}(a), we plot the antenna gains in dB scale for $N_t= 4$, $16$, and $32$ using \eqref{eq:total_antenna_gain_down} and $\phi_{\rm d}=6^\circ$. As expected, the antenna gain increases by $3$ dB for doubling the antenna elements and at the same time, the main beam becomes narrower. To study the impact of this phenomenon, we use the proposed optimal HRA method to calculate the optimal UT angles in USFs for different $N_t$ and report the finding in Fig.~\ref{fig:impact_element}(b). Since antenna with low $N_t$ provides lower gain, the SIRs corresponding to $N_t=4$ obtains lower values. For instance, about $20\%$ of the UAVs suffer from very low SIR (less than $-5$~dB). 

For the other two $N_t$ plots, we can see an interesting trade-off. When $N_t=16$ is considered, Fig.~\ref{fig:impact_hd}(b) verifies that it provides better minimum SIR (greater than $0$~dB) than $N_t=8$, thanks to its higher antenna gain. However, due to its wider beam width, with $N_t=8$, GBSs can cover a larger area in the sky with higher gains. This translates into the fact that about $70\%$ of the UAVs achieve a higher SIR compared to the case when GBSs are equipped with $16$ antennas each. This interesting insight can help the network operators better plan the number of antenna elements they need depending on their performance requirements.

\subsection{Impact of the physical separation of the antenna sets}
\begin{figure}[!t]
\centering{\includegraphics[width=\linewidth]{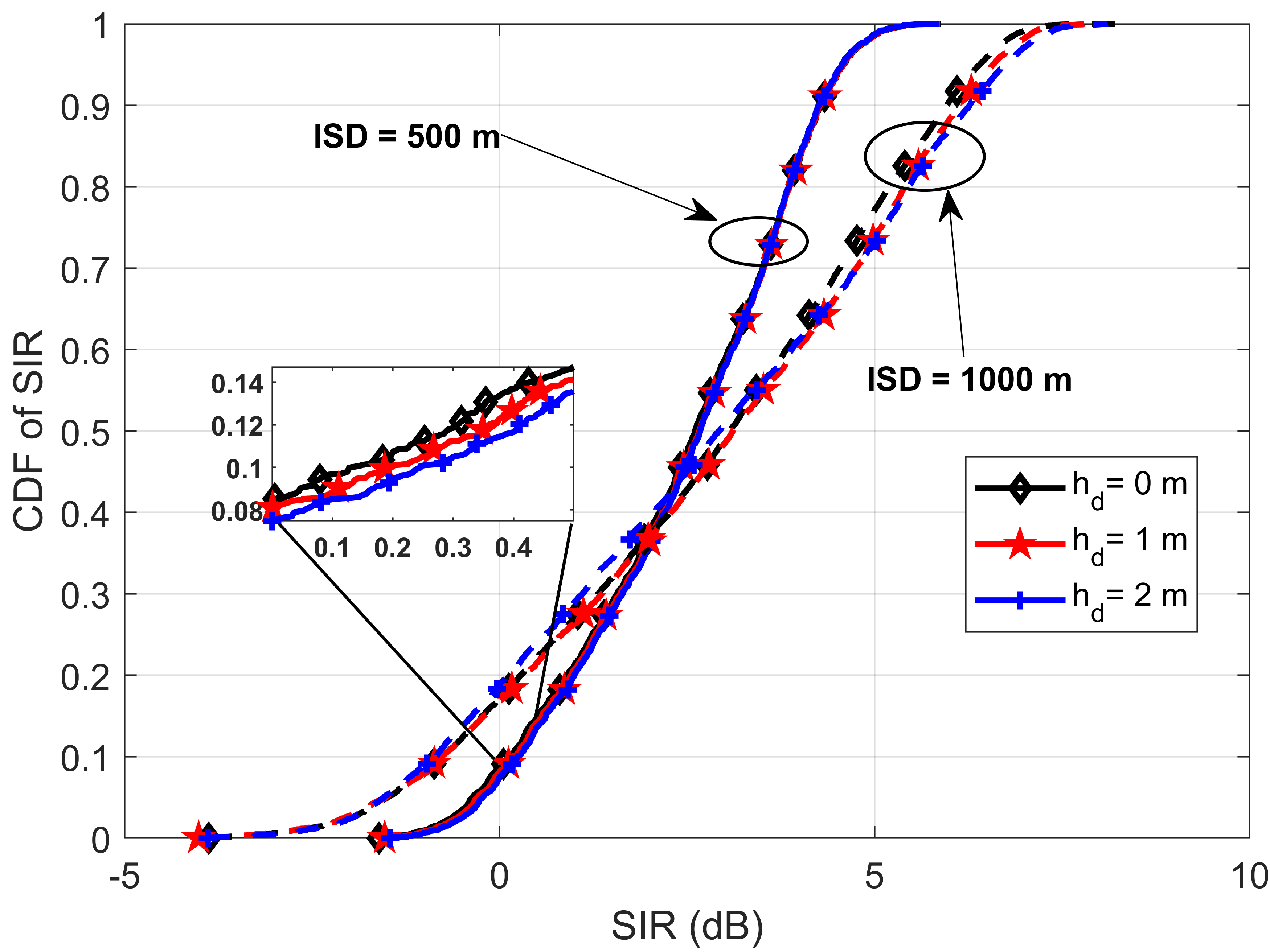}}
    \caption{UAV SIR CDFs for ISD~$=500$~m and ISD~$=1000$~m while considering $h_{\textrm{UAV}}=100$~m. Solid lines represent the SIR with up-tilted antennas and dashed lines represent the case without up-tilted antennas during the USFs. Both of these lines overlap with each other.}
    \label{fig:impact_hd}
\end{figure}


We also study the impact of the antenna separation distance $h_{\rm d}$ between the up-tilted and the down-tilted antenna sets. We consider $h_{\rm UAV}=100$~m and ISD$=500$~m and $1000$~m and show the resulting UAV SIRs for the optimal UT angles in Fig.~\ref{fig:impact_hd}. For both ISDs, we can conclude that the overall impact of $h_{\rm d}$ is very trivial for the optimal UT angles during USFs. The related SIRs are slightly better for $h_{\rm d}=2$~m. 
This is due to the fact that with higher $h_{\rm d}$, the main lobes of the two sets of antennas are more separated from each other and thus creates less interference. 

Another interesting finding is that the impact of $h_{\rm d}$ is more visible for ISD$=1000$~m. This is because the GBSs tend to pick lower UT angles for covering the cell-edge UAVs for larger ISDs, and hence, the higher $h_{\rm d}$ helps to keep the main beams of the up-tilted and down-tilted antennas further away. This results in lower interference and thus higher SIRs for the UAVs. Whereas for lower ISDs, the GBSs pick higher values of UT angles which are already separated from the main beams of the down-tilted antennas, and thus the overall impact of $h_{\rm d}$ is trivial here.

\begin{figure}[!t]
\centering
		\subfloat[ISD $500$~m.]{
			\includegraphics[width=\linewidth]{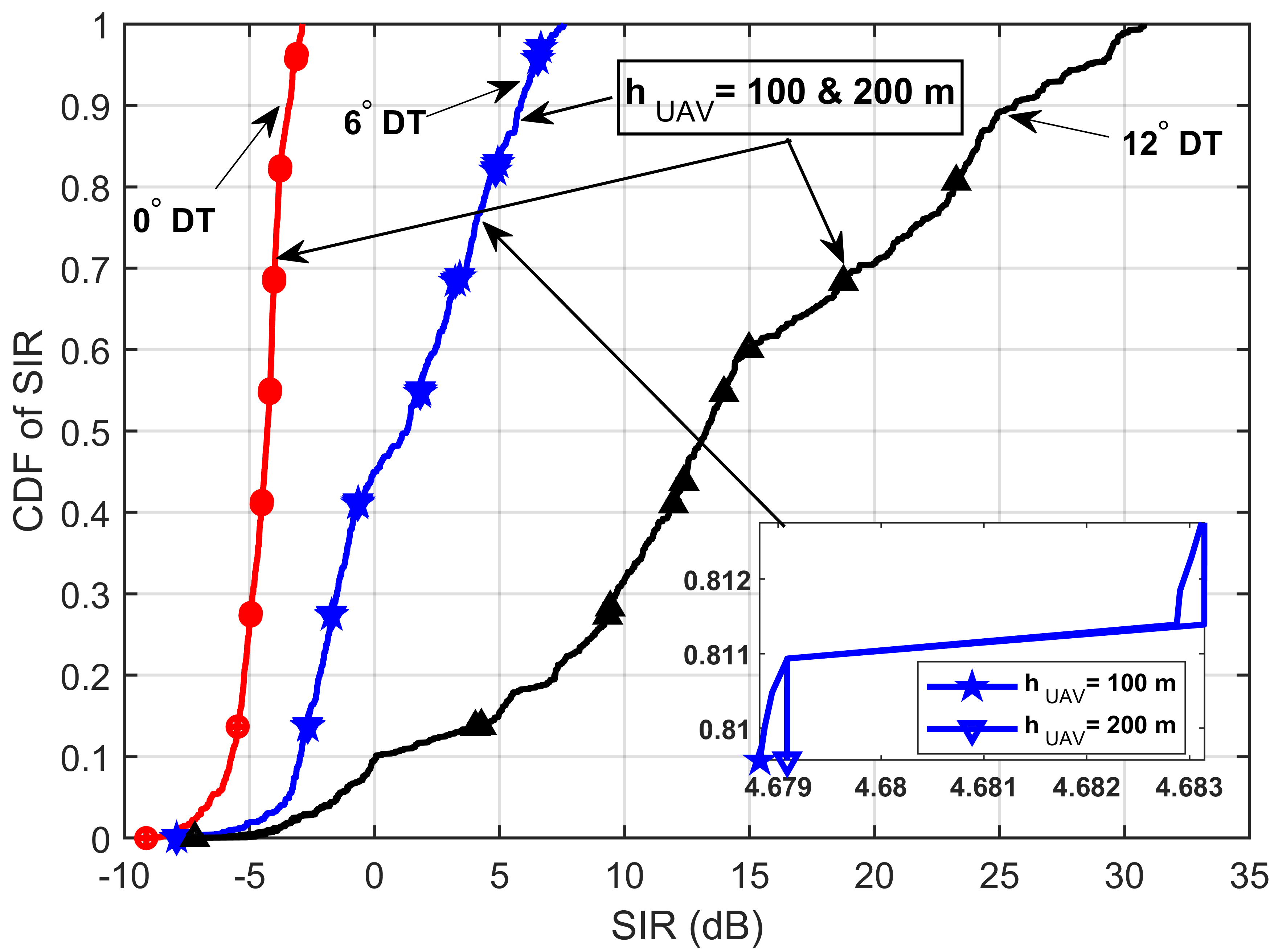}} \hfill
		\subfloat[ISD $1000$~m.]{
			\includegraphics[width=\linewidth]{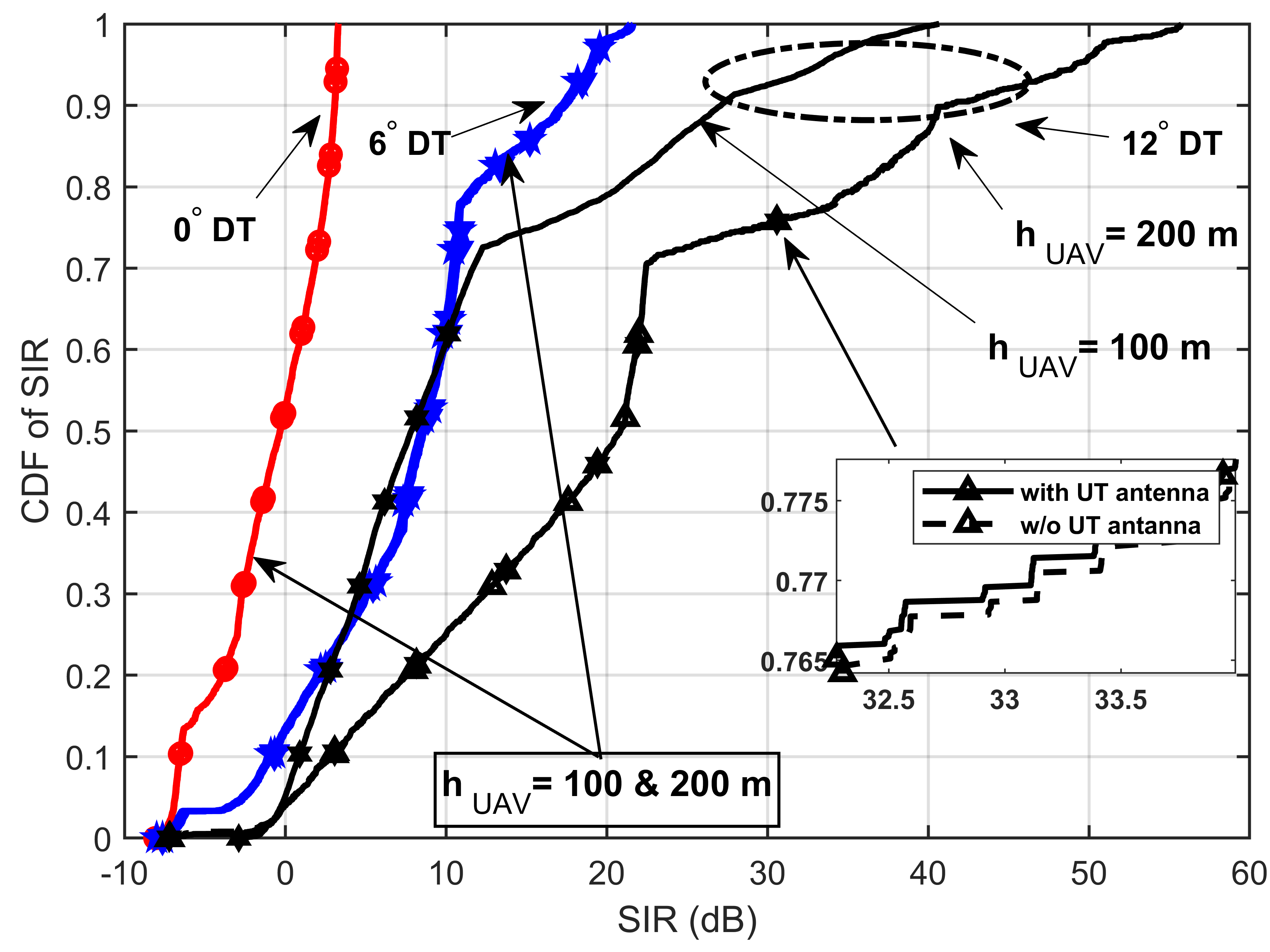}}
		\caption {
		{GUE SIR CDFs with height $1.5$~m for (a) ISD~$=500$~m and (b) ISD~$=1000$~m during USFs. }}
		\label{fig:impact_GUE_rate}
\end{figure}

\subsection{Impact on the GUE SIR}
Thus far, we have focused on scenarios in which the UAVs as the only users in the network. After proper tuning of the UT angles, the presence of the extra set of up-tilted antennas along with the eICIC method can provide a high and reliable SIR for the UAVs flying in the sky. However, the extra set of antennas can also introduce interference to the existing GUEs. Hence, in this subsection, we study the impact of our proposed UT angle tuning scheme on the GUEs.

Here, we consider the three DT angles as done before along with the two ISDs and UAV heights to check the impact thoroughly and report the results in Fig.~\ref{fig:impact_GUE_rate}. We use the GR-based path-loss model with a height of $1.5$~m to represent the GUE cases. We only report the USF results for visual convenience and the CSF cases show the same trends and hence, are omitted here. The cases with the up-tilted antennas are presented with solid lines and scenarios without the up-tilted antennas are represented by the dashed lines. It is evident from the plots of both Fig.~\ref{fig:impact_GUE_rate}(a) and Fig.~\ref{fig:impact_GUE_rate}(b) that the impact of up-tilted antennas on the GUE SIRs is trivial and the lines representing these two scenarios overlap each other. This is because the main lobes of the up-tilted antennas are focused towards the sky and hence, the only impact they can create is through the side lobes. However, these side lobes of the up-tilted antennas can create little to no impact on the GUEs who are associated with GBS providing very high antenna gains. Note that the overall trends will still be the same for 3GPP-based path-loss models~\cite{3gpp.38.901} for GUEs. 

Note that the SIRs of the GUEs increase with increasing DT angle since higher DT angles reduce inter-cell interference. Moreover, larger cell areas or ISDs provide better SIR performance due to the reduced interference on the cell-centered GUEs. Other than the plot for ISD$=1000$~m and $h_{\textrm{UAV}}=200$~m, all other plots show that GUE performance is invariant of the optimal UT angles of the up-tilted antennas. For ISD$=1000$~m and $h_{\textrm{UAV}}=200$~m, the cell-edge users suffer from less interference since GBSs tend to focus more upwards with higher $h_{\textrm{UAV}}$.
\section{Concluding Remarks and Discussion}
\label{sec:Conc}
In this paper, we have proposed a novel cellular architecture by considering an extra set of antennas that are up-tilted to provide good and reliable connectivity to the UAVs. These antennas coexist with the traditional down-tilted antennas and use the same time and frequency resources. The down-tilted antennas can create interference to the UAVs by the antenna side lobes and GR, and we have proposed a modified path-loss model to capture the impact of the GR on the UAVs. To ensure high SIR and reliable connectivity, we have formulated an optimization problem with an aim to maximize the minimum UAV SIR by tuning the UT angle of each GBS. Since the problem is NP-hard, we have proposed a GA-based UT angle optimization method to obtain high-quality suboptimal solutions efficiently. Apart from this, we have also considered the 3GPP specified eICIC to reduce the interference caused by the down-tilted antennas. We have run extensive simulations to study our proposed method for various cellular network deployment configurations such as ISD, UAV height, DT angle, number of antenna elements, etc. Our results have shown that overall our proposed method can provide high minimum SIR for the UAVs. Our results have also revealed some interesting design guidelines such as the impact of the number of antenna elements and the DT angles on the UAV SIR performance, and most importantly, our method has shown little to no impact on the SIRs of the existing GUEs in the network. Thus, the proposed technique can be a strong candidate for deploying large-scale urban aerial systems in the near future while maintaining the reliable and efficient coexistence of UAVs and GUEs.

Our proposed framework can be extended in several ways. First of all, the duty cycle parameter $\beta$ can be taken into account in the optimization framework to maximize the minimum rate (instead of SIR) of both GUE and UAV since those who are associated with down-tilted antennas suffer from the reduced rate in our proposed framework. Moreover, the updated version of eICIC known as further enhanced ICIC (FeICIC) can be considered in which traffic data is transmitted during ABS with relatively low power. Another interesting study will be providing better connectivity and reliable mobility (i.e., reducing ping-pong and handover failures) to the UAVs whose trajectories are known before. It is worth noting that, our proposed method will not be able to support UAVs in the regions where cellular infrastructures are not available i.e., over deserts or oceans. We may need to rely on high-altitude aerials platforms or low earth orbital satellites for providing reliable connectivity to UAVs in these extreme cases.

Another limitation of our proposed framework is that the extra set of antennas will increase the overall energy consumption of the network. Moreover, the DT angles of the down-tilted antennas can impact the SIR performance of the UAVs. Hence, joint optimization of UT angles, transmit power of the up-tilted antennas, eICIC/FeICIC parameters, and DT angles will be included in our future work to make our framework more efficient.

\bibliographystyle{myIEEEtran} 
\bibliography{reference}


\end{document}